%% file: EFT_over_the_edge.tex
\documentclass[amsmath,amssymb,aps,prd,11pt,tightenlines,superscriptaddress,
nofootinbib,preprintnumbers,notitlepage]{revtex4-1}
\usepackage{multirow}
\usepackage{graphicx}
\usepackage{tabularx}
\usepackage{slashed}
\usepackage{url}
\usepackage{hyperref}
\usepackage[maxfloats=100]{morefloats}
\usepackage{color,xcolor}
\usepackage{booktabs}
\usepackage[normalem]{ulem}

\makeatletter
\def\l@subsubsection#1#2{}
\makeatother

\input{declare}

\AtBeginDocument{
  \heavyrulewidth=.08em
  \lightrulewidth=.05em
  \cmidrulewidth=.03em
  \belowrulesep=.65ex
  \belowbottomsep=0pt
  \aboverulesep=.4ex
  \abovetopsep=0pt
  \cmidrulesep=\doublerulesep
  \cmidrulekern=.5em
  \defaultaddspace= .5em
  \setlength{\tabcolsep}{0.5em}
}

\begin{document}


\title{Pushing Higgs Effective Theory over the Edge} 

\author{Anke Biek\"otter}
\affiliation{Institut f\"ur Theoretische Teilchenphysik und Kosmologie, RWTH Aachen, Germany}

\author{Johann Brehmer}
\affiliation{Institut f\"ur Theoretische Physik, Universit\"at Heidelberg, Germany}

\author{Tilman Plehn}
\affiliation{Institut f\"ur Theoretische Physik, Universit\"at Heidelberg, Germany}

\date{\today}

\begin{abstract} 
  Based on a vector triplet model we study a possible failure of
  dimension-6 operators in describing LHC Higgs kinematics. First, we
  illustrate that including dimension-6 contributions squared can significantly
  improve the agreement between the full model and the dimension-6
  approximation, both in associated Higgs production and in
  weak-boson-fusion Higgs production. Second, we test how a simplified
  model with an additional heavy scalar could improve the agreement in
  critical LHC observables.  In weak boson fusion we find an
  improvement for virtuality-related observables at large energies,
  but at the cost of sizeable deviations in interference patterns and
  angular correlations.
\end{abstract}

\maketitle
\tableofcontents

\clearpage

\section{Introduction}
\label{sec:intro}

After the discovery of a light Higgs boson~\cite{higgs,discovery}, one
of the key tasks of the LHC is to test if the observed particle indeed
corresponds to the minimalist setup of the Higgs sector in the
Standard Model. Because of the many intricacies of the electroweak
sector of the Standard Model, it is not straightforward to define a
theoretical framework which describes possible deviations in the Higgs
sector. If we want to remain more general than testing specific
models~\cite{bsmreview}, we can use an effective field theory ansatz.
Here the Lagrangian is organized by the field or particle content, the
symmetry structure, and the mass
dimension~\cite{eftfoundations,eftorig,higgsreview}.

The problem at least for the interpretation of Run~I results in terms
of an effective theory is the limited experimental accuracy. Combined
with an assumed maximum size of underlying couplings the experimental
precision determines the maximum testable hierarchy of
scales~\cite{legacy,heft_limitations}.  If for weakly interacting new
physics models we assume that the higher-dimensional operators are
ordered by factors $g^2 m_h^2/\Lambda^2$, or that the only relevant
scale of Higgs production is given by $m_h$, a typical LHC accuracy of
$10\%$ on a rate measurement translates into a new physics reach
around
\begin{align}
\left| \frac{\sigma \times \text{BR}}{\left( \sigma \times \text{BR} \right)_\text{SM}} - 1 \right|
= \frac{g^2 m_h^2}{\Lambda^2} \gtrsim 10\%
\qquad \Leftrightarrow \qquad 
\Lambda < \frac{g \, m_h}{\sqrt{10\%} } \, \approx 400~\gev \; .
\label{eq:lmax}
\end{align}
We assume $g <1$, implying a reasonably weakly interacting theory,
equivalent to Wilson coefficients of order one. This is exactly the
$\Lambda$ range we find in the full analysis~\cite{legacy}.

This does \emph{not} mean that an analysis of LHC data in
terms of a truncated dimension-6 Lagrangian cannot be useful, but it
does require us to carefully check the correspondence between the
dimension-6 Lagrangian and complete models. It turns out that a
dimension-6 Lagrangian describes weakly interacting extensions of the
Higgs-gauge sector at the LHC well~\cite{too_long}.  One key
ingredient to this success is a $v$-improved matching
procedure~\cite{too_long}, which includes effects of the Higgs VEV in
the matching to a Lagrangian with linearly realized electroweak
symmetry breaking. The simplest realization of this idea is to set the
new physics scale $\Lambda$ to the physical mass of new particles
including contributions from the Higgs VEV instead of using the mass
scale in the unbroken phase of the electroweak symmetry. This
effectively absorbs effects from dimension-8 operators into the
dimension-6 Lagrangian by replacing $\phi^\dagger \phi \to v^2 /2
$. These terms are negligible when $v \ll \Lambda$, but can
significantly improve the accuracy of the dimension-6 model in
scenarios without a clear separation of scales~\cite{too_long}.

This study rests on the findings of the more theoretical discussion in
Ref.~\cite{too_long}. For models where that discussion does not lead
to an obvious conclusion it attempts to answer two experimental key
questions:
\begin{enumerate}
\item Is it justified or preferable to include dimension-6 operators
  squared while neglecting dimension-8 operators interfering with the
  Standard Model?
\item How can we improve our description when even the $v$-improved
  matching starts to fail and new states affect the LHC kinematics?
\end{enumerate}
The first point is of immediate practical relevance for fits of Wilson
coefficients to LHC data and has been discussed from different
perspectives~\cite{legacy,too_long,mvh,gino,square_dim_6_others,spanno}.
\bigskip

These questions cannot be answered without making assumptions about
the new physics scenario. We will illustrate them in a specific setup
with a modified gauge sector~\cite{gauge_modifications}, a setup known
to challenge the dimension-6 framework~\cite{too_long}.  We will rely
on an extension of the Standard Model by a massive vector field
$\tilde{V}^a_\mu$ which is a triplet under $SU(2)_L$ and has the mass
$M_{\tilde{V}}$~\cite{anke, too_long}.\footnote{Such a model is not
  manifestly renormalizable, but can be embedded in a UV-complete
  theory in which $V$ acquires its mass through a Higgs
  mechanism~\cite{gauge_modifications}.} Its Lagrangian includes the
terms
\begin{alignat}{5}
\lag \supset& \,
  - \dfrac{1}{4}\,\tilde{V}_{\mu\nu}^a\,\tilde{V}^{\mu\nu\, a}
  + \dfrac{M_{\tilde{V}}^2}{2}\,\tilde{V}_\mu^a\,\tilde{V}^{\mu\,a}
  + i\,\frac{g_V}{2} \,c_H\,\tilde{V}_\mu^a\,\left[\phi^\dagger \sigma^a \,\overleftrightarrow{D}^\mu\,\phi\,\right]
  +\dfrac{g_w^2}{2 g_V}\,\tilde{V}_\mu^a\,\sum_\text{SM fermions}\, c_F \overline{F}_L\,\gamma^\mu\, \sigma^a \,F_L
 \notag \\
 &+ \dfrac{g_V}{2}\,c_{VVV}\,\epsilon_{abc}\,\tilde{V}_\mu^a\,\tilde{V}_\nu^b\,D^{[\mu}\tilde{V}^{\nu]c}
  + g_V^2\,c_{VVHH}\,\tilde{V}_\mu^a\,\tilde{V}^{\mu a}\,(\pbp) \,
  - \dfrac{g_w}{2}\,c_{VVW}\,\epsilon_{abc}\,W^{\mu\nu}\,\tilde{V}_\mu^b\,\tilde{V}_\nu^c \,.
 \label{eq:lag-vectortriplet}
\end{alignat}
Five coupling parameters $c_j$ describe the different interactions
of the new vector triplet to itself, the Standard Model fermions $F$,
the Higgs doublet $\phi$, and the $SU(2)_L$ field strength $W^{\mu
  \nu}$.  The covariant derivative acts on the triplet as $D_\mu\,
\tilde{V}_\nu^a = \partial_\mu\, \tilde{V}_\nu^a+g_V \epsilon^{abc}\,
\tilde{V}^b_\mu \tilde{V}_\nu^c$.  The coupling constant $g_V$ is the
characteristic strength of the heavy vector-mediated interactions,
while $g_w$ denotes the $SU(2)_L$ gauge coupling. After mixing they
will combine to the observed weak gauge coupling. This mixing of the
new heavy states with the weak bosons combined with the new, heavy
resonances is what can lead to large effects at the LHC.

\begin{table}[t]
\begin{tabular}{lll} 
  \toprule
  \multicolumn{3}{c}{HISZ basis} \\
  \midrule
  $\ope{\phi1} = (D_\mu\phi)^\dagger \, \phi\,\phi^\dagger \, (D^\mu\phi)$  &
  $\ope{\phi2} = \dfrac{1}{2}\partial^\mu(\phi^\dagger\phi)\,\partial_\mu(\phi^\dagger\phi)$ &
  $\ope{\phi3} = \dfrac{1}{3}(\pbp)^3$ \\[4mm]
  $\ope{GG} = (\pbp)\,G^A_{\mu\nu}\,G^{\mu\nu\, A}$ \\[2mm]
  $\ope{BB} = -\dfrac{g'^2}{4}(\pbp)\,B_{\mu\nu}\,B^{\mu\nu}$ &
  $\ope{WW} = -\dfrac{g^2}{4}(\pbp)\,W^k_{\mu\nu}\,W^{\mu\nu\, k}$ &
  $\ope{BW} = -\dfrac{g\,g'}{4}(\phi^\dagger\sigma^k\phi)\,B_{\mu\nu}\,W^{\mu\nu\, k}$ \\[4mm]
  $\ope{B}  = \dfrac{ig}{2}(D^\mu\phi^\dagger)(D^\nu\phi)\,B_{\mu\nu}$ &
  $\ope{W} = \dfrac{ig}{2}(D^\mu\phi^\dagger)\sigma^k( D^\nu\phi)\,W_{\mu\nu}^k$ \\[4mm]
  $\ope{u\phi} =(\phi^\dagger\phi)(\bar Q_3 \tilde \phi u_{R})$ &
  $\ope{d\phi} =(\phi^\dagger\phi)(\bar Q_3 \phi d_{R})$ &
  $\ope{e\phi} =(\phi^\dagger\phi)(\bar L_3 \phi e_{R})$ \\[2mm]
  \bottomrule
\end{tabular}
\caption{Bosonic CP-conserving Higgs operators in the HISZ basis.}
\label{tab:operators}
\end{table}

Unlike in Ref.~\cite{too_long} we now define our dimension-6
Lagrangian in the HISZ basis~\cite{hisz}, to make our results
compatible with the \textsc{SFitter} Run~I legacy analysis of
Ref.~\cite{legacy}:
\begin{equation}
  \lag \supset \sum_i \frac {f_i} {\Lambda^2} \, \ope{i} \,,
\end{equation}
with the Higgs operators $\ope{i}$ defined in
Tab.~\ref{tab:operators}. Our $v$-improved matching scale is $\Lambda
= m_{\xi}$, the mass of the neutral heavy particle $\xi^0$ after mixing
of the new $V$ state with the $Z$ boson. The dimension-6 Wilson
coefficients for the triplet model read
\begin{align}
  f_{\phi 2} &= \frac 3 4 \left( -2 \,c_F \, g^2 + c_H \, g_V^2 \right) \,, \quad&\quad f_{WW} &= c_F \, c_H \notag \\
  f_{\phi 3} &= -3 \lambda \left( -2 \, c_F \,  g^2 + c_H \, g_V^2 \right) \,,  \quad&\quad f_{BW} &= c_F \, c_H \equiv f_{WW} \notag  \\
  f_{f \phi} &= - \frac 1 4 \, y_f \, c_H \left( -2 \, c_F \, g^2 + c_H \, g_V^2 \right)  \,, \quad&\quad f_W &= - 2 \, c_F \,c_H \; .
\end{align}

Structurally, those dimension-6 operators are of two types,
\begin{align}
\mathcal{O} \propto \frac{g^2 v^2}{\Lambda^2}
\qquad \text{and} \qquad 
\mathcal{O} \propto \frac{g^2 \partial^2}{\Lambda^2} \; .
\label{eq:operators}
\end{align}
The second type of operator, such as $\ope{WW}$, introduces a momentum
dependence in the $VVh$ interaction and thereby modifies kinematic
distributions in Higgs-strahlung and weak-boson-fusion (WBF) Higgs
production.  Problems with our $v$-improved dimension-6 approximation
for LHC kinematics typically occur through those operators. For two
benchmark points introduced in
Tab~\label{tab:triplet_benchmarks}~\cite{too_long}, which challenge
the agreement between full model and dimension-6 approximation, we
give the Wilson coefficients in Tab.~\ref{tab:benchmarks}. The
benchmark point T1 features constructive interference between the
dimension-6 amplitudes and the SM in weak boson fusion and destructive
interference in $Vh$ production, while for T4 it is the other way
around. In this paper we will again study two particularly sensitive
observables, the $m_{Vh}$ (or $p_{T,V}$) distribution in
Higgs-strahlung and the $p_{T,j}$ distribution in weak boson
fusion. In models with extended Higgs sectors the $m_{hh}$
distribution in Higgs pair production would develop very similar
features, including an $s$-channel resonance, but is experimentally
less pressing~\cite{hh,too_long}.

\begin{table}[b!]
  \renewcommand{\arraystretch}{1.2}
  \setlength{\tabcolsep}{0.5em}
  \centering
    \begin{tabular}{c c rrrrrr  c rrrrr}
      \toprule
      \multirow{2}{*}{Benchmark} &\hspace*{1em}& \multicolumn{6}{c}{Triplet model} &\hspace*{1em}&  \multicolumn{5}{c}{Dimension-6 approximation}  \\
      \cmidrule{3-8} \cmidrule{10-14}
      && $M_{\tilde{V}}$ & $g_V$ & $c_H$ & ${c}_{F}$ & ${c}_{VVHH}$  & $m_\xi$ && $f_{\phi 2}$ & $f_{\phi 3}$ & $f_{WW}$ & $f_W$ & $f_{u \phi 33}$ \\
      \midrule
      T1 && 591 & 3.0 & $-0.47$ & $-5.0$ & 2.0 & 1200 && 0.00 & 0.00 & 2.45 & $-4.90$ & 0.00 \\
      T4 && 1246 & 3.0 & $-0.50$ & $3.0$ & $-0.2$ & 1200 && 2.64 & $-1.37$ & $-1.56$ & 3.12 & $-0.87$ \\
      \bottomrule
    \end{tabular}
    \caption{Benchmark points for the vector triplet model and
      matched Wilson coefficients for the dimension-6 model.
      All masses are given in GeV. Table from Ref.~\cite{too_long}. }
  \label{tab:benchmarks}
\end{table}

\section{To square or not to square}
\label{sec:square_dim6}

If we accept that a dimension-6 Lagrangian describing Higgs signatures
at the LHC is not necessarily part of a consistent effective field
theory, but rather a successful and reproducible parametrization of
weakly interacting new physics, there exists no fundamental
motivation~\cite{legacy,too_long,mvh,gino,square_dim_6_others,spanno}
to include or to not include the dimension-6 squared term in
\begin{align}
|\mat_{4+6}|^2 = |\mat_4|^2 + 2 \, \text{Re} \mat_4^* \mat_6 \stackrel{?}{+} |\mat_6|^2 \; .
\end{align}
A dimension-6 squared term of comparable or larger size than
the interference term can appear in phase-space regions with
a suppressed dimension-4 prediction, even when the EFT expansion in
$E/\Lambda$ holds and dimension-8 effects are negligible.
In the absence of any first-principle reason how to treat this term,
we need to test the different possibilities from a practical
perspective.

\subsection*{Higgs-strahlung}

\begin{figure}[t]
  \centering
  \includegraphics[width=\textwidth]{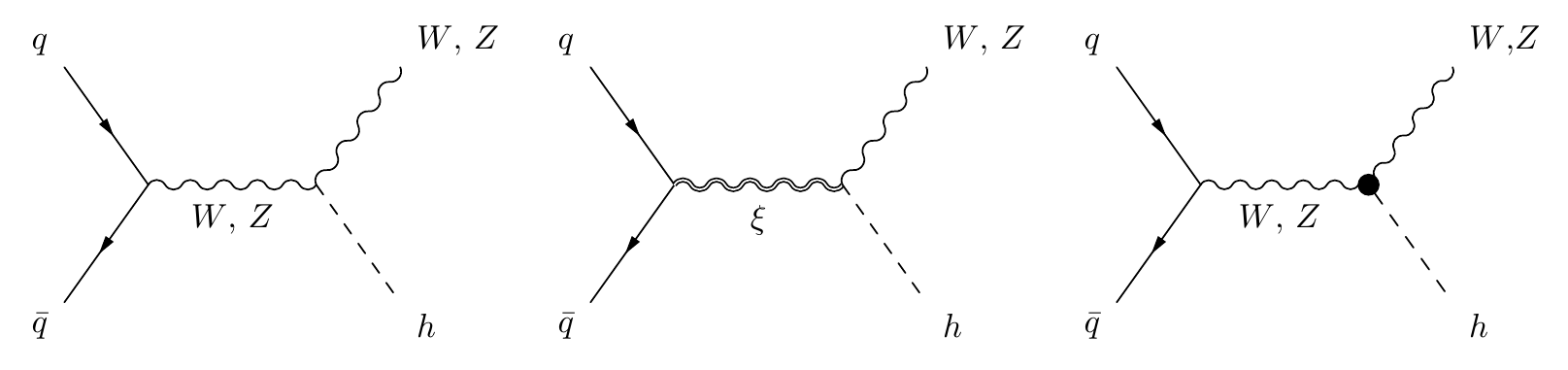}
  \caption{Example diagrams for $Vh$ production in the SM (left), in
    the vector triplet model (middle), and in the EFT (right), where
    the blob denotes effects from the dimension-6 operators.}
  \label{fig:diag_VH}
\end{figure}

\begin{figure}[t]
  \includegraphics[width=0.43\textwidth]{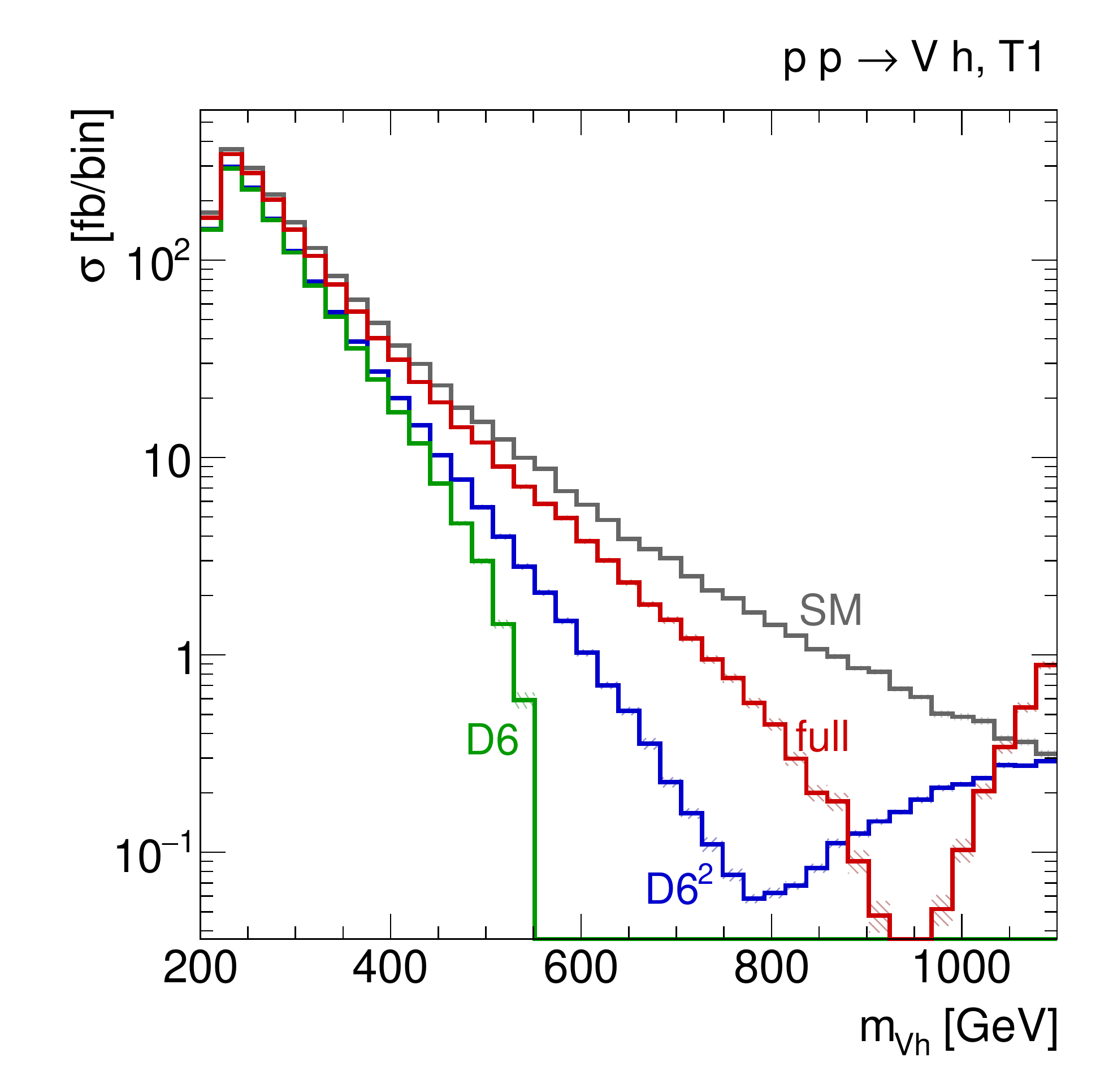}
  \hspace*{0.05\textwidth}
  \includegraphics[width=0.43\textwidth]{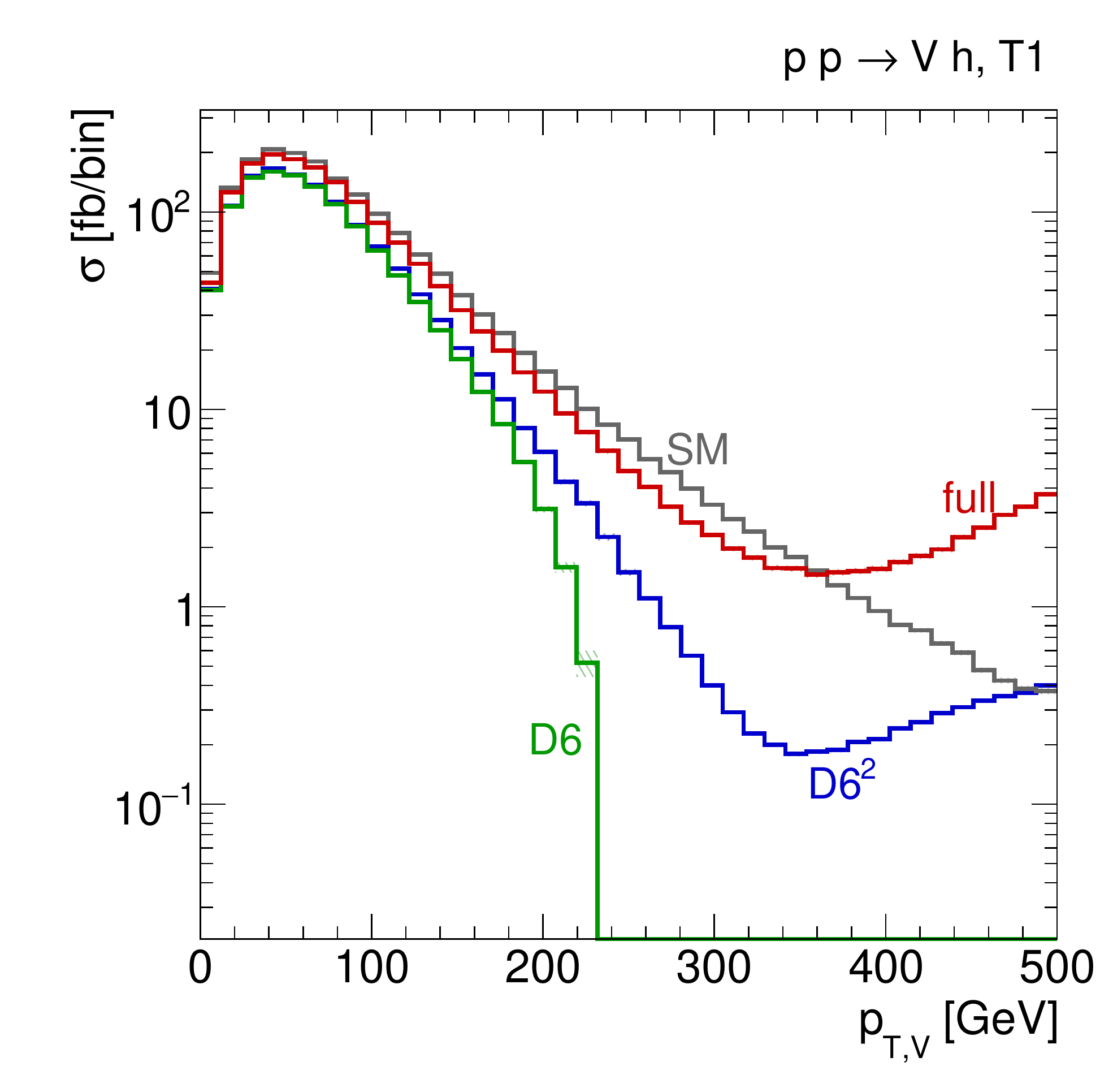} \\
  \includegraphics[width=0.43\textwidth]{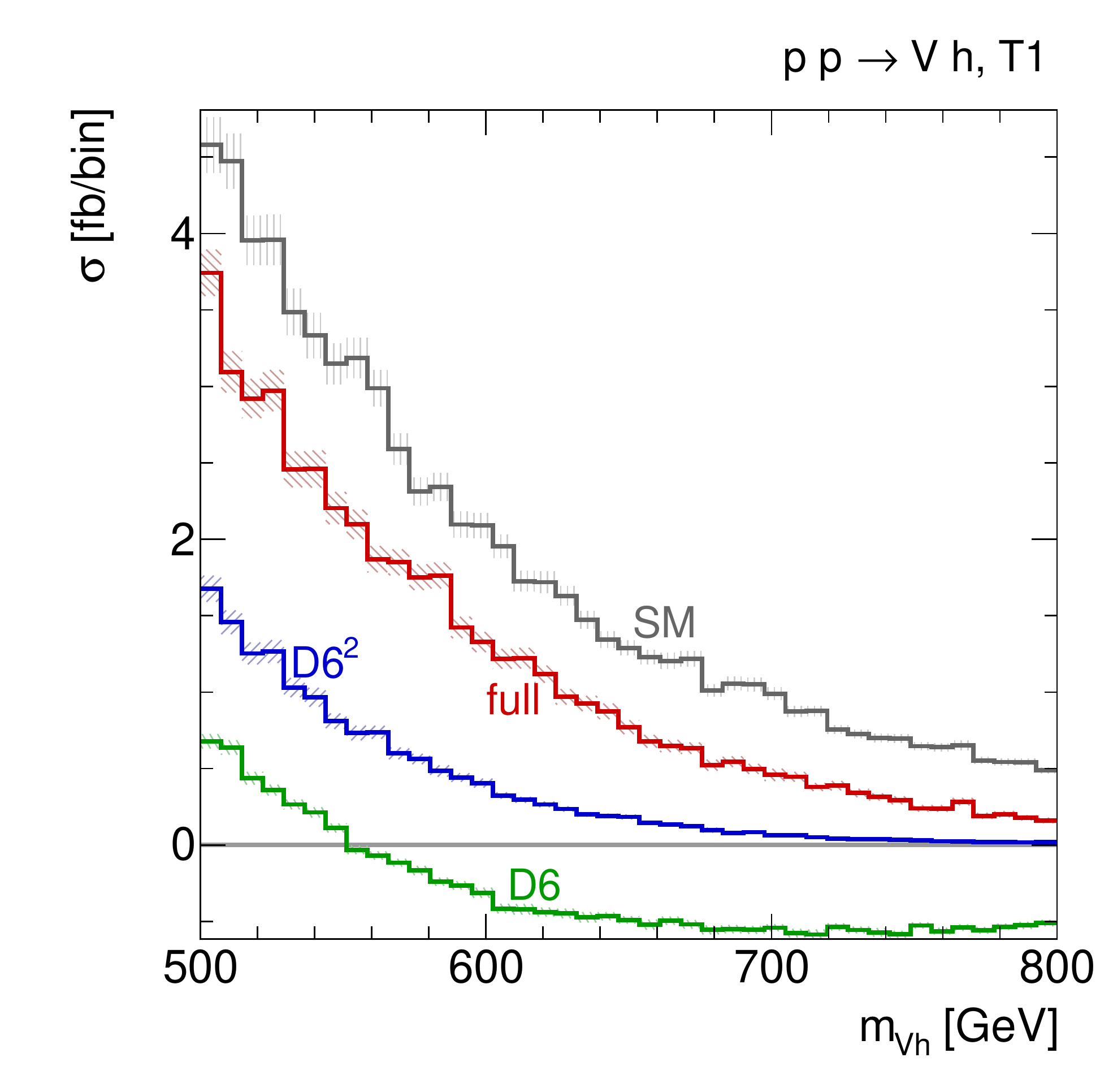} 
  \hspace*{0.05\textwidth}
  \includegraphics[width=0.43\textwidth]{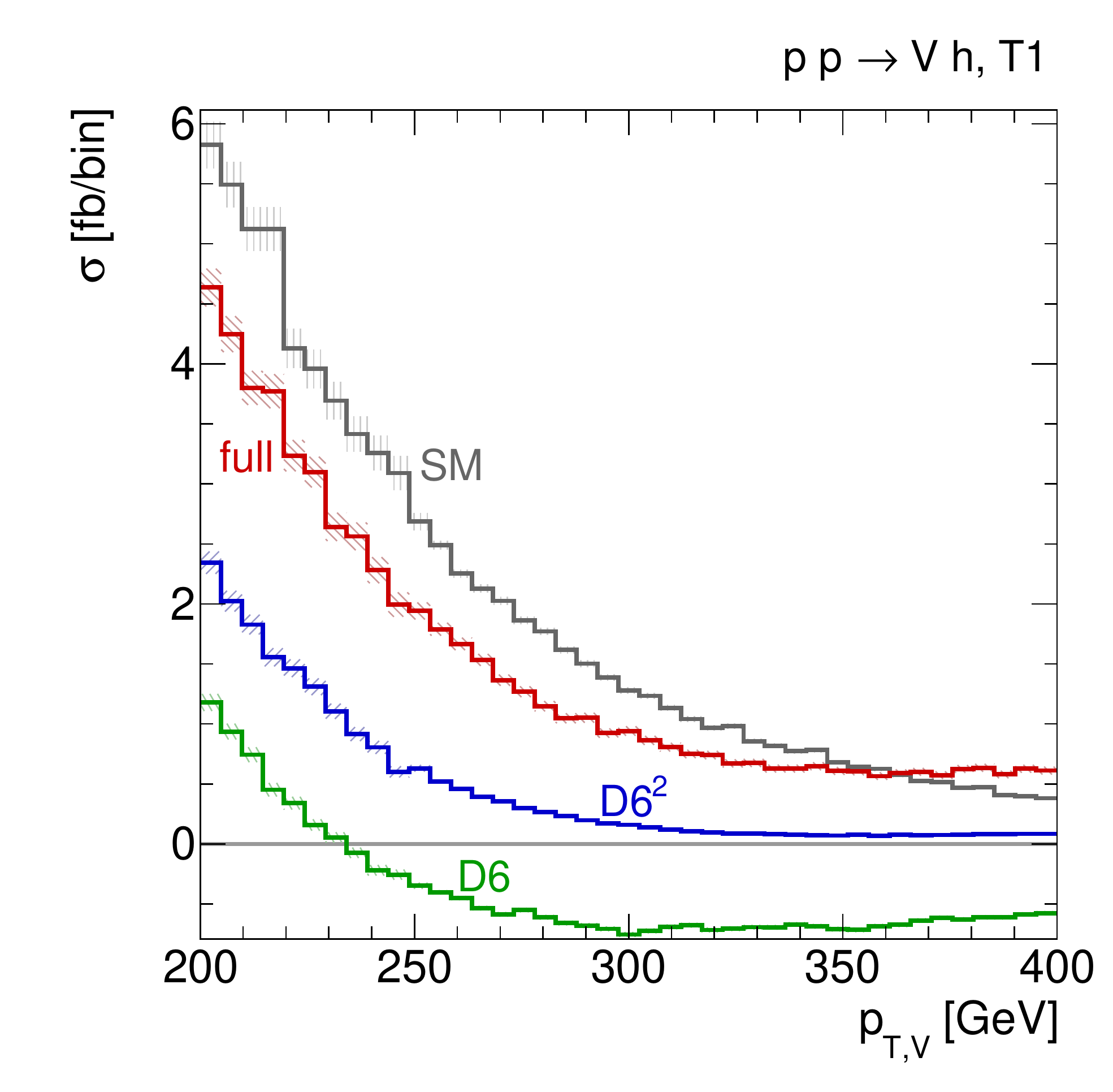} \\
  \includegraphics[width=0.43\textwidth]{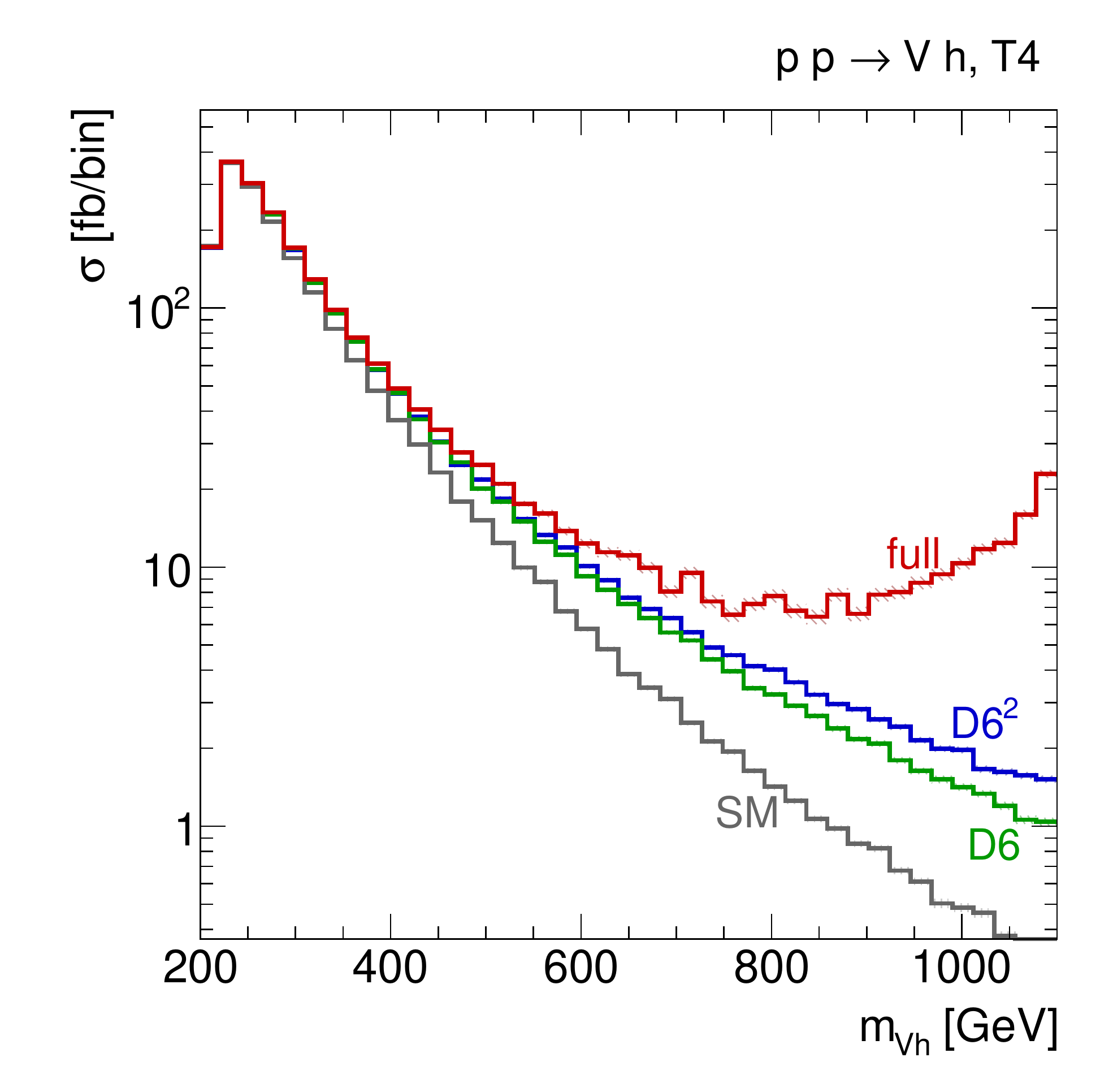}
  \hspace*{0.05\textwidth}
  \includegraphics[width=0.43\textwidth]{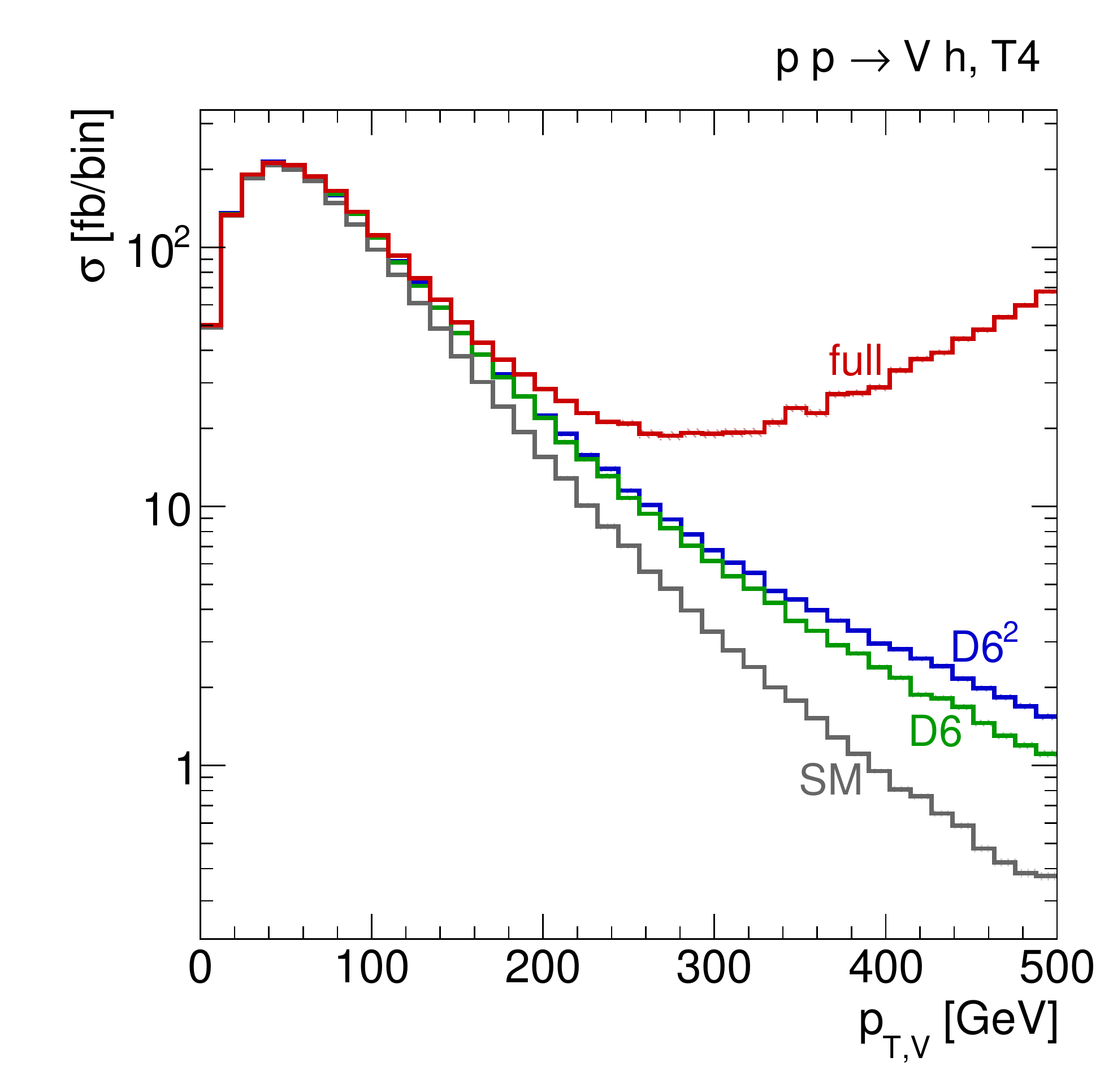}
  \caption{$Vh$ distributions with (``D6$^{2}$'') and without (``D6'')
    the dimension-6 squared term. The left panels show $m_{Vh}$, the right panels
    $p_{T,V}$. The central panels show the region where leaving out
    the squared dimension-6 terms leads to a negative cross section.}
  \label{fig:squared_VH}
\end{figure}

We first analyze associated $Vh$ production with $V =
W^\pm, Z$ at 13~TeV LHC energy. To retain as much phase space as
possible we only consider the parton-level process
\begin{align}
  p p \to V h
\end{align}
simulated in \toolfont{MadGraph}~\cite{madgraph} without cuts or
decays, see Fig.~\ref{fig:diag_VH}. It is easy to see where in phase
space the effective theory breaks down: for on-shell outgoing Higgs
and gauge bosons a large momentum flow through the Higgs operator can
only be generated through the virtual $s$-channel propagator. We can
directly test this in the observable $m_{Vh}$ distribution, comparing
the full model with the dimension-6 approach at large momentum flow.

We show the $m_{Vh}$ distributions in the left panels of
Fig.~\ref{fig:squared_VH}.  While theoretically the $m_{Vh}$
distribution is cleaner, for example when we include initial state
radiation, we can see the same effects in the highly correlated
$p_{T,V}$ distribution (right panels), due to the simple $2 \to 2$
signal kinematics~\cite{mvh,gino}.  The T1 benchmark point is
constructed with a low new physics scale and a destructive
interference between Standard Model and dimension-6 term. We see that
the squared dimension-6 terms are clearly needed to avoid negative
cross sections in the high-energy tails of the distributions. Driven
by the light new particles, inconsistencies otherwise occur around
\begin{align} 
m_{Vh} > 600~\gev \approx \frac{m_\xi^\text{(T1)} }2
\qquad \text{or} \qquad 
p_{T,V} > 300~\gev \approx \frac{m_\xi^\text{(T1)} }{4} \; ,
\label{eq:breakdown_vh}
\end{align}
clearly within reach of Run~II. The reason why differences appear much
below $m_{Vh} = m_\xi$ is that the new states are wide and their pole
contribution extends through a large interference effect. Because for
this benchmark point the discrepancies signal the onset of a new
$s$-channel propagator pole, the agreement between full model and
dimension-6 operators is limited and will hardly improve once we
include for example dimension-8 terms~\cite{kilian}.\footnote{We are 
  convinced that, 
  if the LHC experiments should observe such a new
  resonance, the justification of a dimension-6 description will most
  likely not be of experimental or theoretical concern.}

For the constructively interfering benchmark point T4 we observe no
dramatic effects in the tails, but the agreement between the full
model and the dimension-6 approximation is improved when we include
these terms. Both benchmark points therefore suggest to include the
dimension-6 squared terms in the LHC analysis, to improve the
agreement between the model and the dimension-6 Lagrangian.

\subsection*{WBF Higgs production}

\begin{figure}[t]
  \centering
  \includegraphics[width=\textwidth]{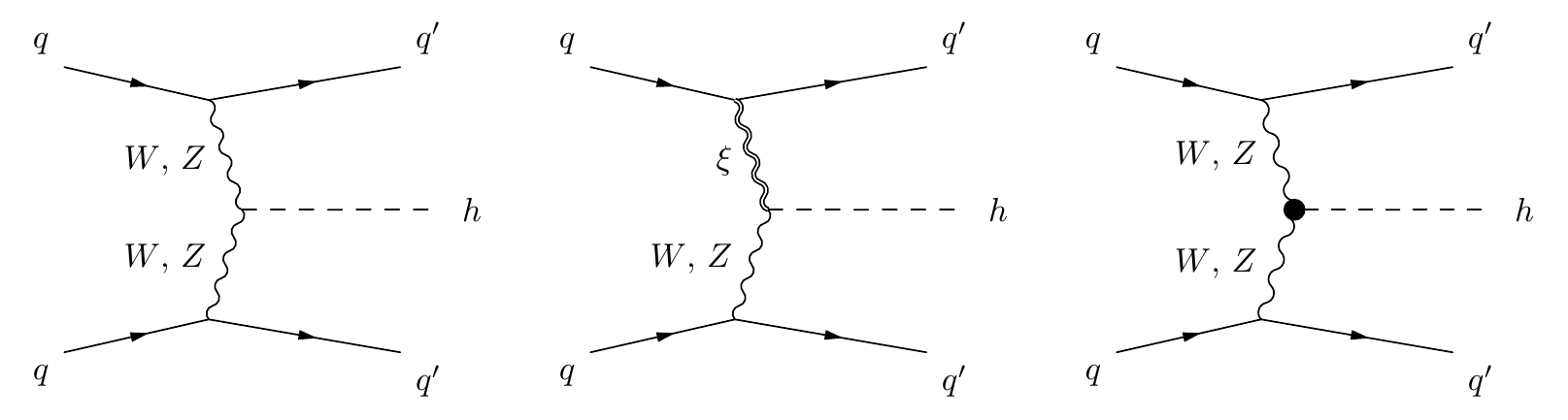}
  \caption{Example diagrams for WBF Higgs production in the SM (left),
    in the vector triplet model (middle), and in the EFT (right),
    where the blob denotes effects from the dimension-6 operators.}
  \label{fig:diag_WBF}
\end{figure}

\begin{figure}
  \includegraphics[width=0.43\textwidth]{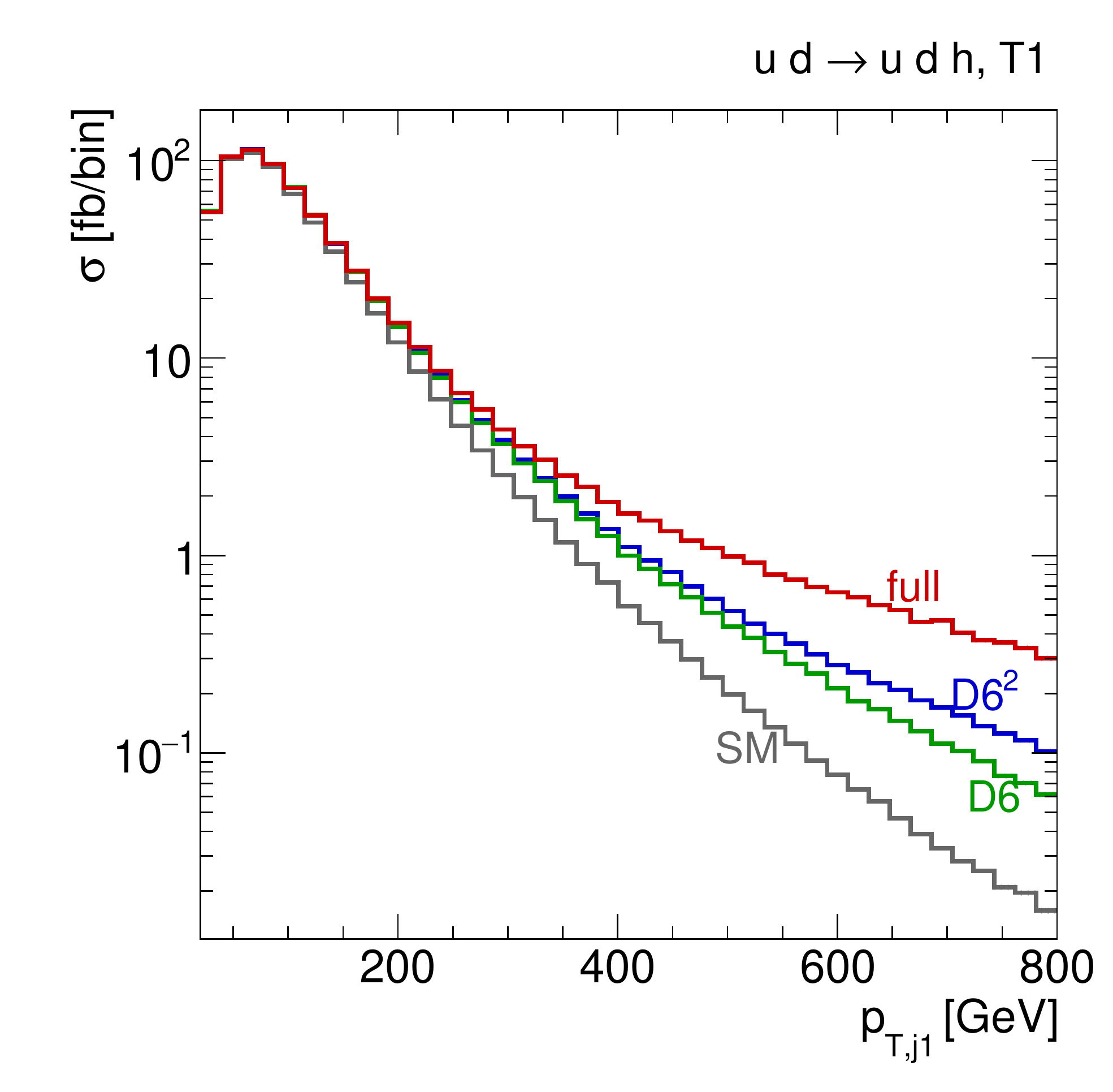} 
  \hspace*{0.05\textwidth}
  \includegraphics[width=0.43\textwidth]{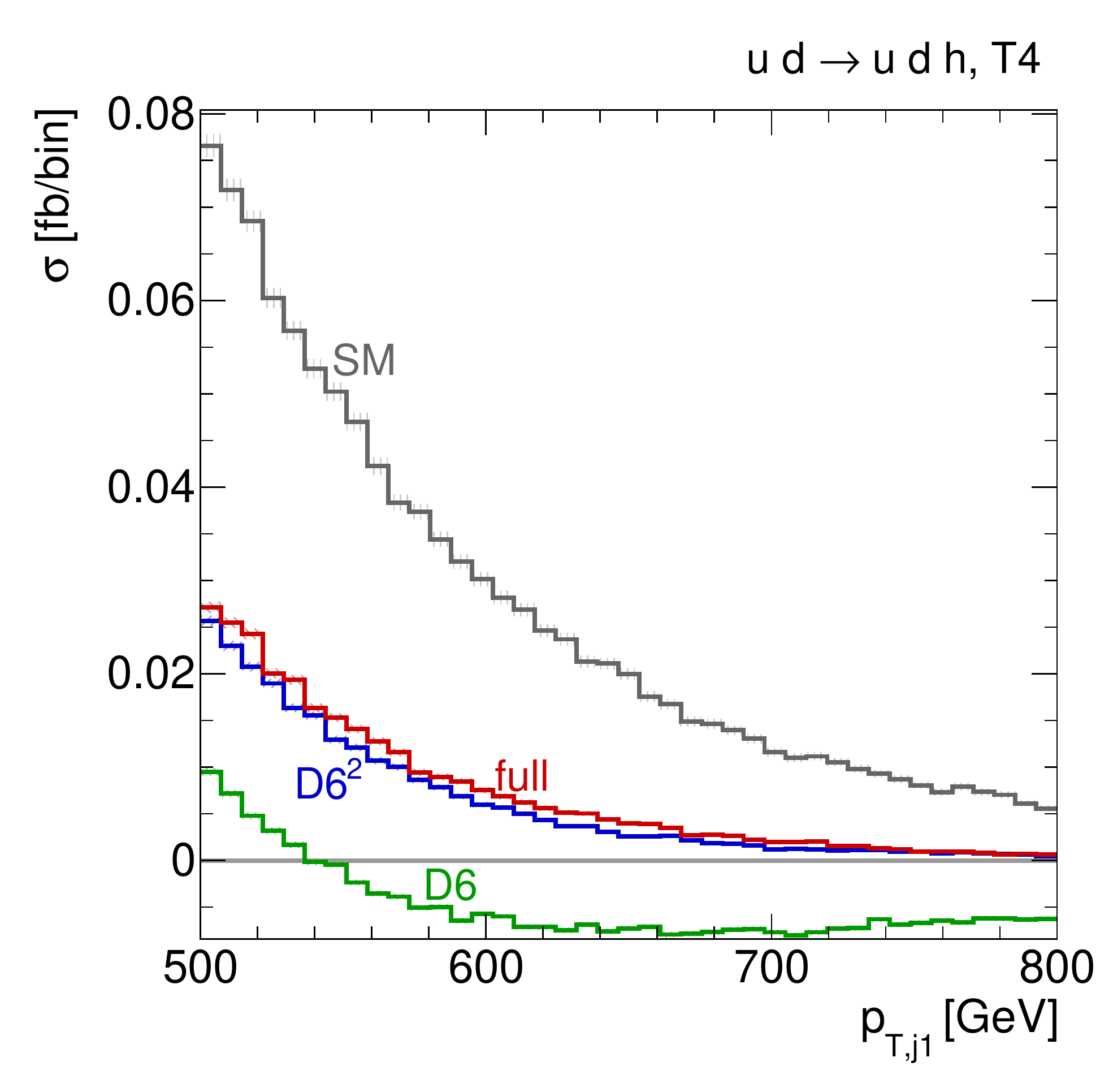}\\
  \includegraphics[width=0.43\textwidth]{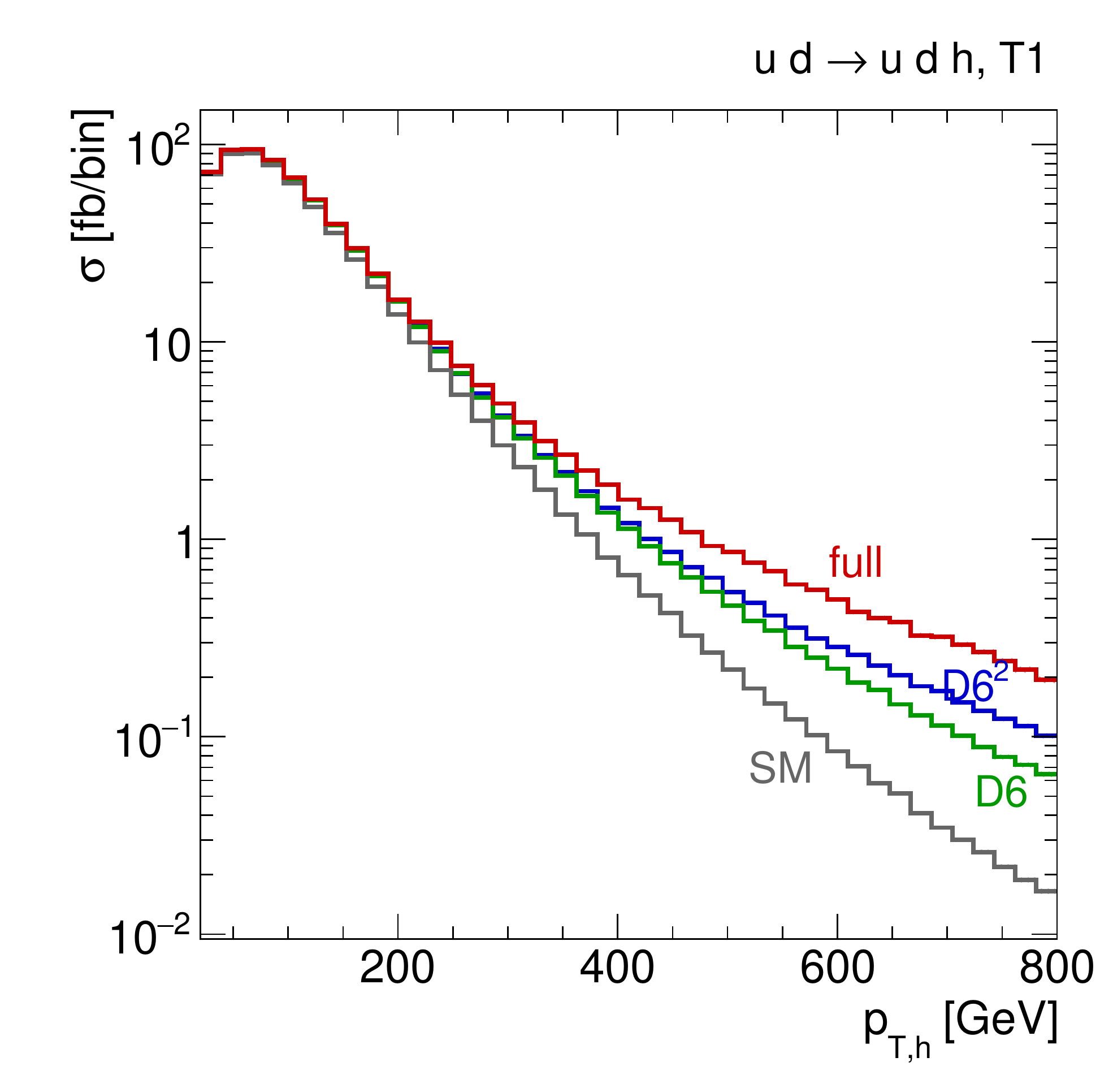} 
  \hspace*{0.05\textwidth}
  \includegraphics[width=0.43\textwidth]{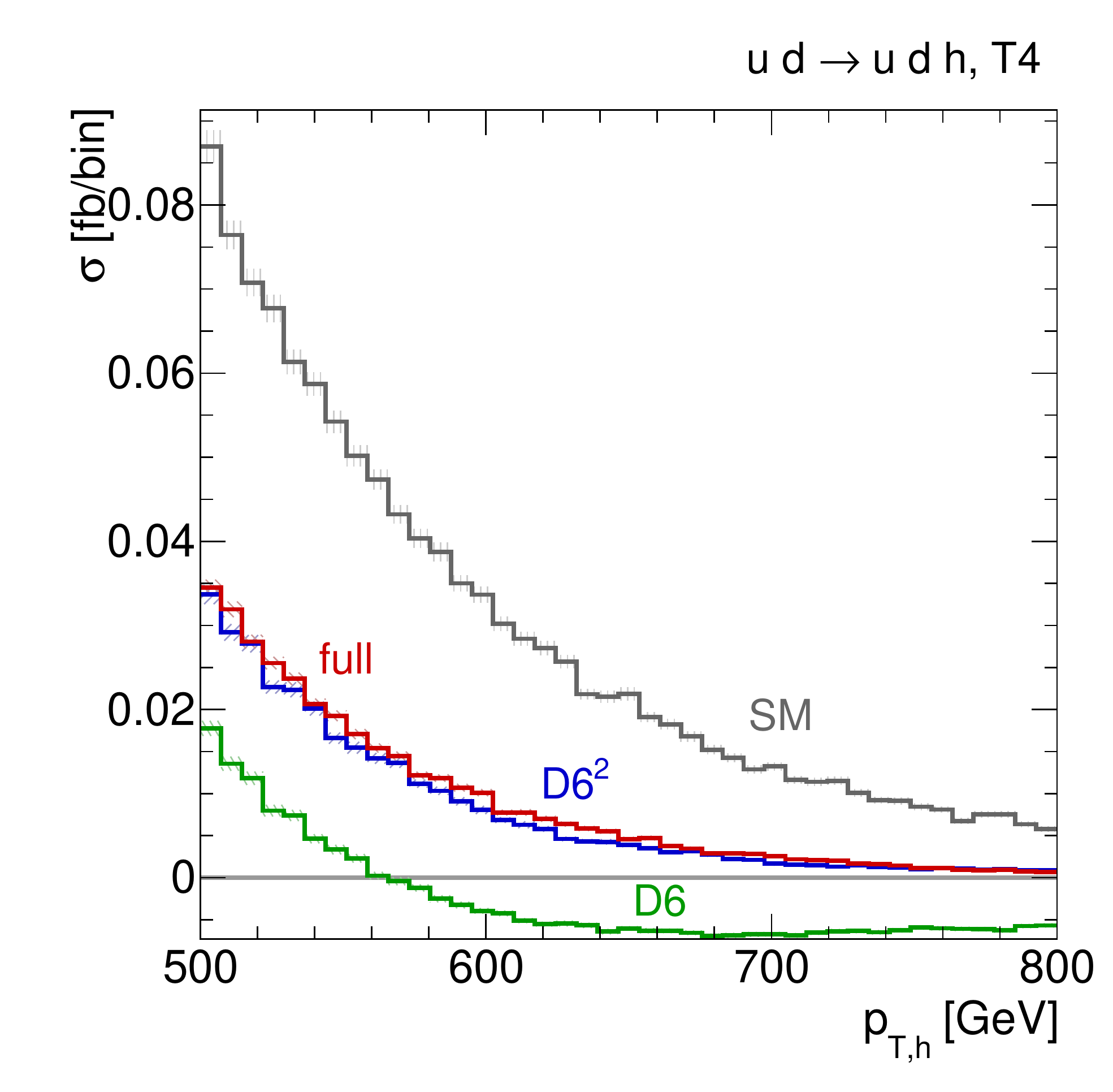}\\
  \includegraphics[width=0.43\textwidth]{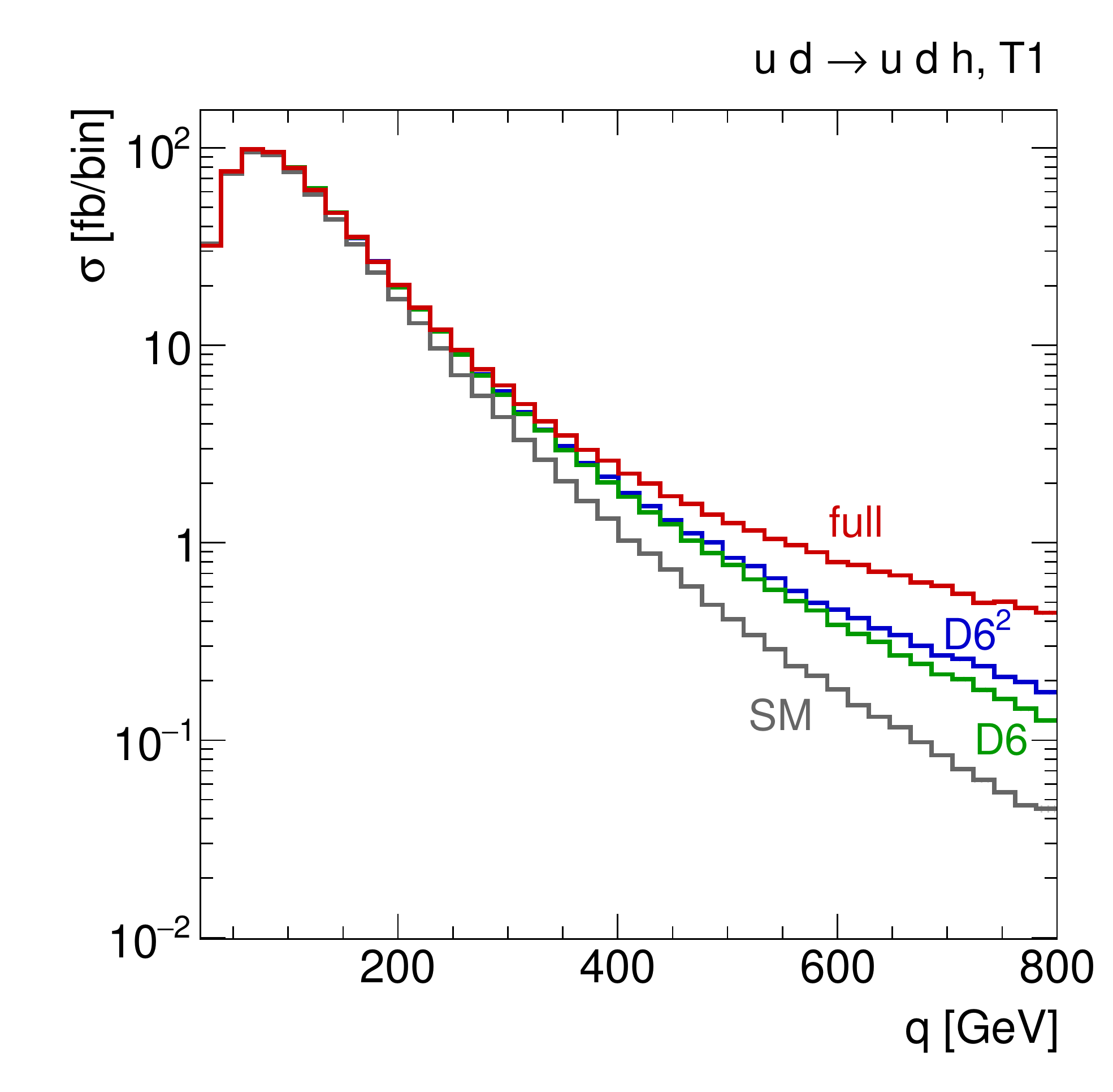} 
  \hspace*{0.05\textwidth}
  \includegraphics[width=0.43\textwidth]{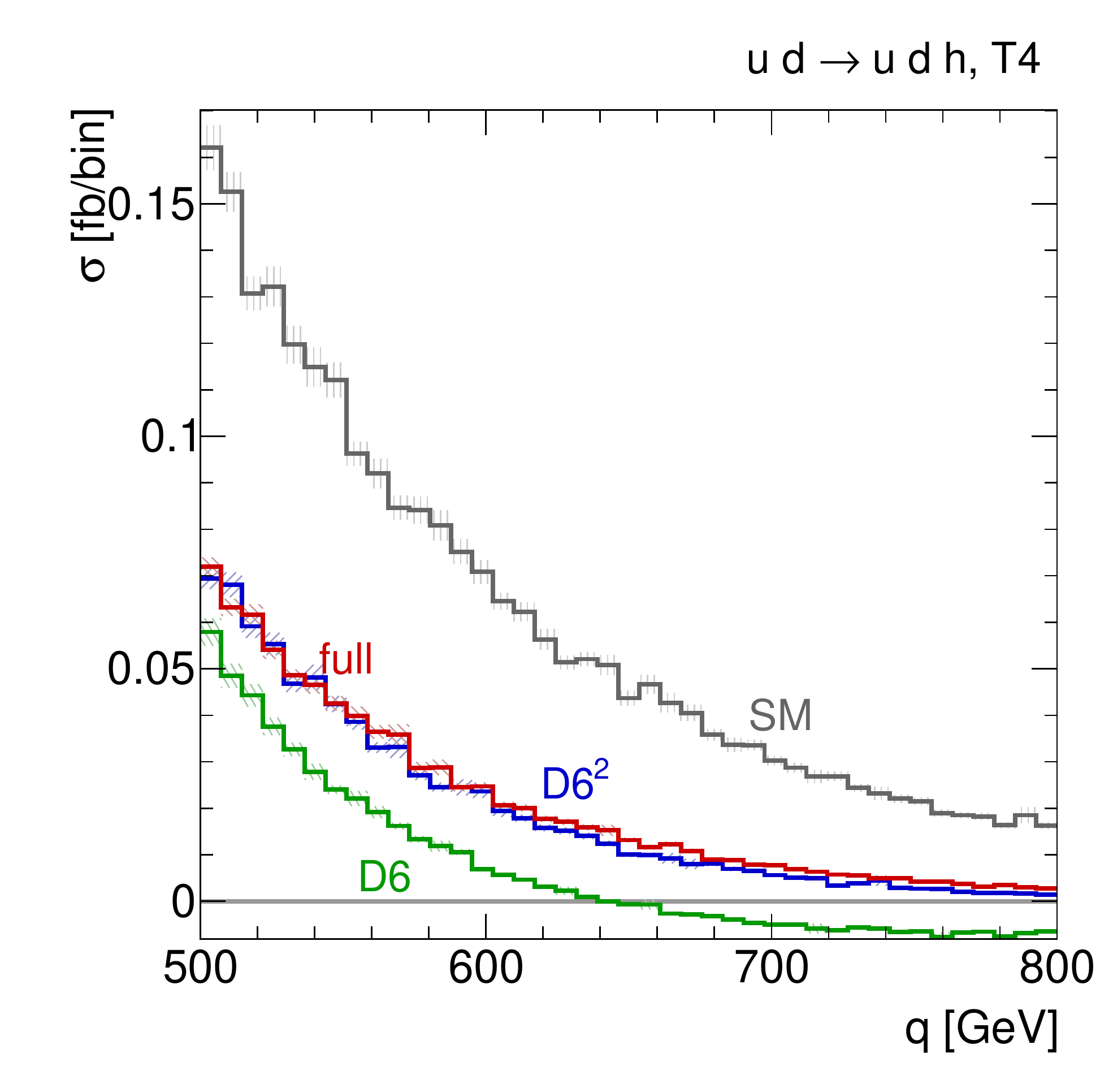}
  \caption{WBF distributions with (``D6$^{2}$'') and without (``D6'') the
    dimension-6 squared term. From top to bottom: $p_{T,j1}$, $p_{T,h}$, and
    virtuality $q$ defined in Eq.\;\eqref{eq:virt}. The right panels show the region where
    leaving out the squared dimension-6 terms leads to a negative cross
    section.}
  \label{fig:squared_WBF}
\end{figure}

Weak-boson-fusion Higgs production is a $2 \to 3$ process with two
$t$-channel gauge bosons carrying the momentum to the Higgs vertex,
see Fig.~\ref{fig:diag_WBF}.  The relevant kinematic variables are the
two virtualities of the weak bosons. Following many studies in the
framework of the effective $W$
approximation~\cite{effective_w,polarized_ww} it is straightforward to
link them to the $p_T$ of the tagging jets, which even for multiple
jet radiation can be linked to the transverse momentum of the
Higgs~\cite{Buschmann:2014twa} (even though it is not clear if this
distribution is theoretically or experimentally favored).  Again, we
start with the parton-level signal process
\begin{align}
u d \to u' d' h
\label{eq:def_wbf}
\end{align}
with only one minimal cut $p_{T,j} > 20$~GeV for the two tagging jets.  We
show the results for the now constructively interfering benchmark
point T1 and the now destructively interfering benchmark point T4 in
Fig.~\ref{fig:squared_WBF}. Negative event rates for T4 appear around
\begin{align}
p_{T,j_1} > 600~\gev \approx \frac{m_\xi^\text{(T4)}}{2} \; , 
\label{eq:breakdown_wbf}
\end{align}
forcing us to either disregard the corresponding model hypothesis or
to add the dimension-6 squared term.  For the less critical point T1
the agreement between the vector triplet model and its dimension-6
approximation including the squared terms extends well into the range
where deviations from the Standard Model become visible.

\begin{figure}[t]
  \includegraphics[width=0.43\textwidth]{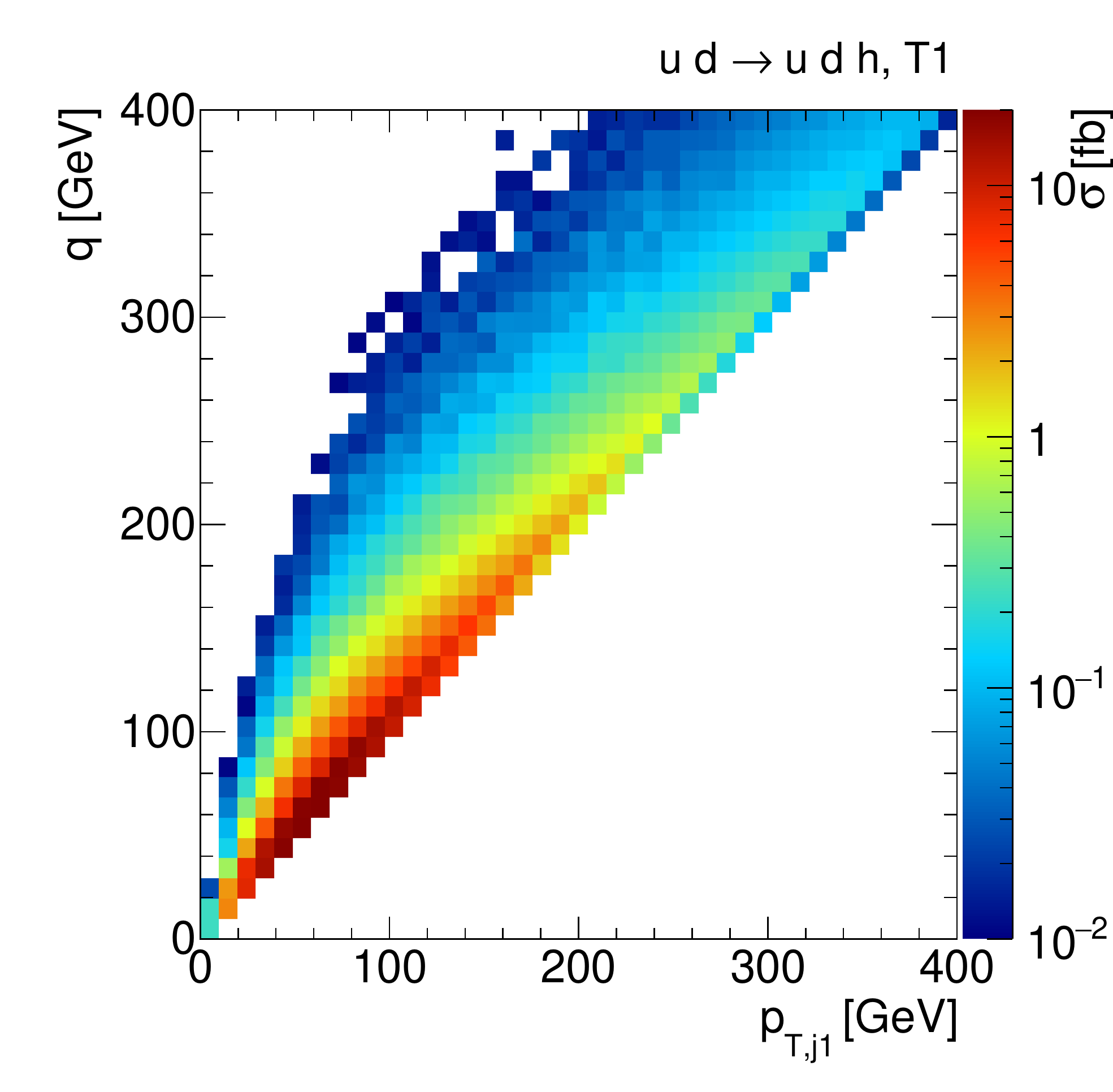}
  \hspace*{0.05\textwidth}
  \includegraphics[width=0.43\textwidth]{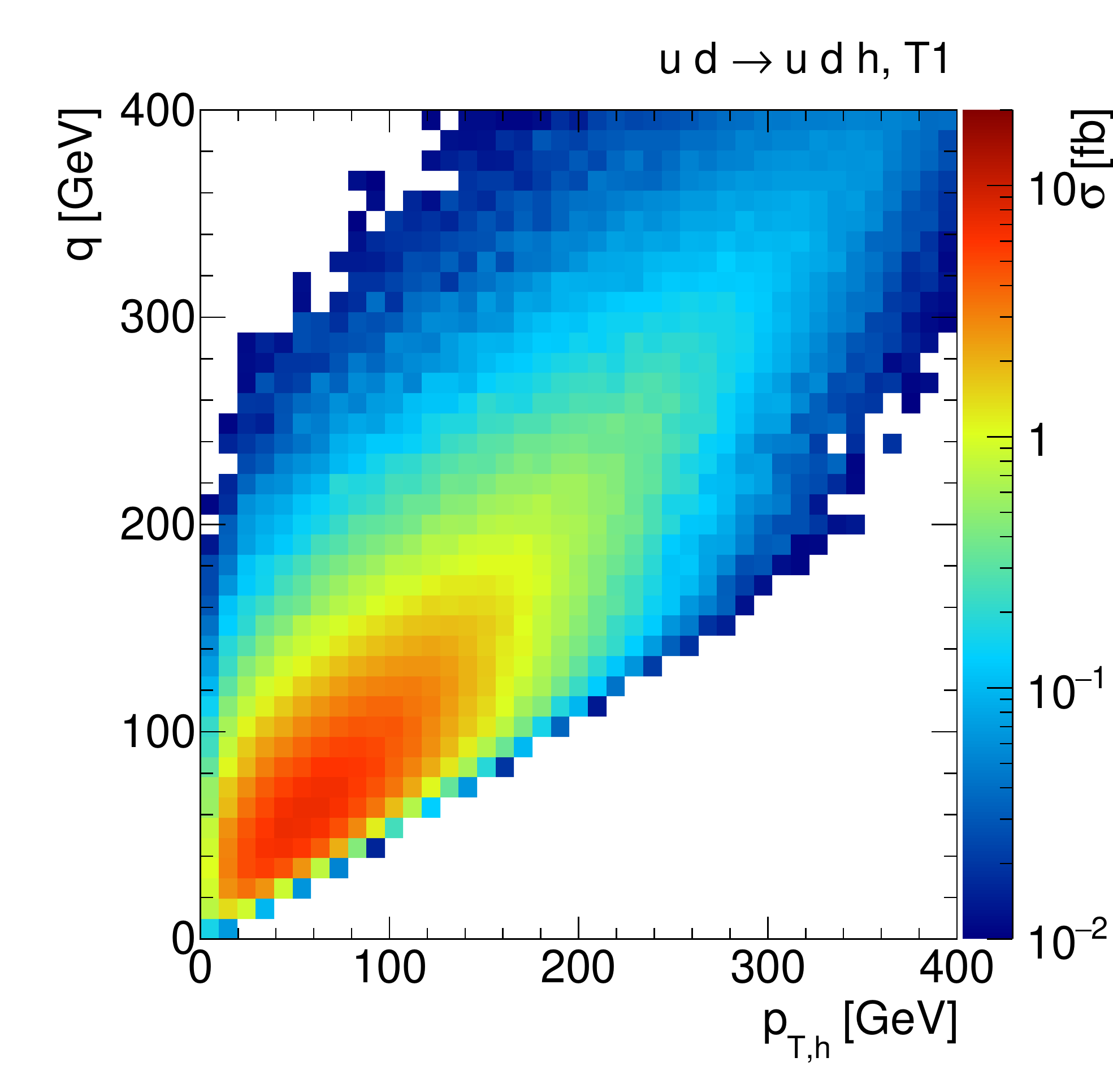} 
  \caption{WBF correlations between the virtuality $q$ and
    $p_{T,j_1}$ (left) or $p_{T,h}$ (right).}
  \label{fig:virt_corr}
\end{figure}

\begin{figure}[b!]
  \includegraphics[width=0.43\textwidth]{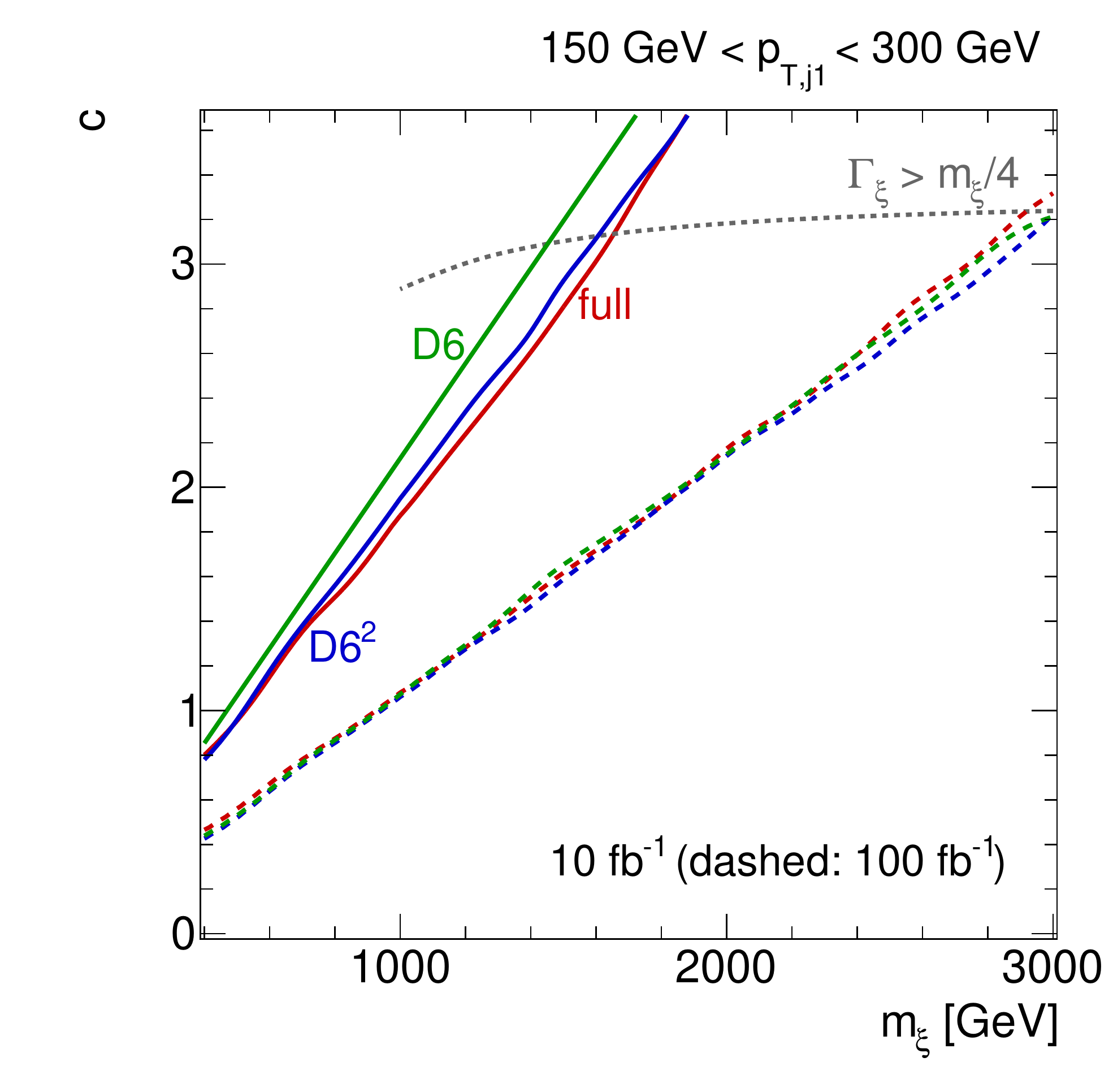}
  \hspace*{0.05\textwidth}
  \includegraphics[width=0.43\textwidth]{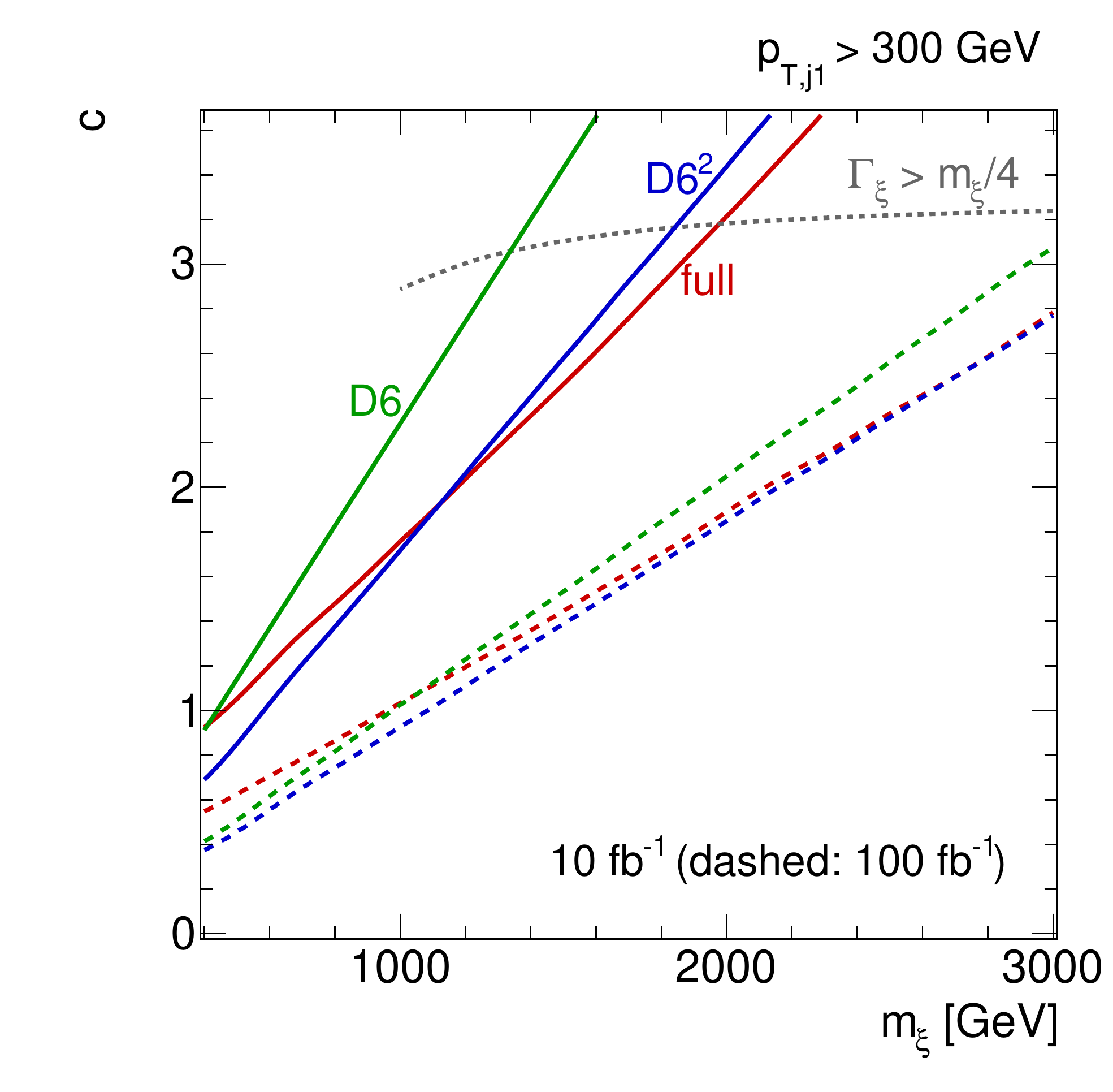} 
  \caption{Expected limits on a two-dimensional slice of the vector
    triplet parameter space. We show the analysis based on the event
    numbers in $150~\gev < p_{T,j_1} < 300~\gev$ (left) and based on
    the tail $p_{T,j_1} > 300$~GeV.}
  \label{fig:limits}
\end{figure}

In the middle panels of Fig.~\ref{fig:squared_WBF} we see that indeed
the $p_{T,h}$ distribution looks almost identical to $p_{T,j_1}$. Both
of them can be traced back to the unobservable virtualities of the
weak bosons. Due to the preferred collinear direction of the
quark-vector splittings, the $W$-mediated and $Z$-mediated diagrams
populate very different parton-level phase-space regions, with
basically no interference between them.  We can thus define the
virtuality variable~\cite{gino,polarized_ww}
\begin{align}
  q =
  \begin{cases}
    \max\left(   \sqrt{ | (p_{u'} - p_{d})^2 | } \, , \, \sqrt{ | (p_{d'} - p_{u})^2 | }  \right) & \text{for $W$-like phase-space points,} \\
    \max\left(   \sqrt{ | (p_{u'} - p_{u})^2 | } \, , \, \sqrt{ | (p_{d'} - p_{d})^2 | }  \right)  & \text{for $Z$-like phase-space points,}
  \end{cases}
  \label{eq:virt}
\end{align}
with the distribution shown in the bottom panels of
Fig.~\ref{fig:squared_WBF}. Comparing it to $p_{T,h}$ and $p_{T,j_1}$ we see
essentially the same behavior.  The strong correlation of $q$ with
the observable transverse momenta of the leading tagging jet and the
Higgs is explicitly shown in Fig.~\ref{fig:virt_corr}.\bigskip

Finally, we compare expected exclusion limits on the vector triplet in
the absence of a signal, based on the full model vs the dimension-6
approach.  For the process shown in Eq.\;\eqref{eq:def_wbf} we
multiply the cross sections with a branching ratio
$\br(h \to 2\ell 2\nu) \approx 0.01$.  We disregard non-Higgs
backgrounds as well as parton-shower or detector effects.  We then
count events in two high-energy bins of the $p_{T,j_1}$ distributions,
defining a parameter point to be excluded if $S/\sqrt{S+B} > 2$.
While this statistical analysis is not designed to be realistic, it
illustrates how the validity of our dimension-6 approach affects
possible limits.  For our limit setting procedure we choose a
two-dimensional plane defined by $m_\xi$ versus a universal coupling
rescaling $c$,
\begin{align}
  g_V = 1 \; , \qqquad 
  c_H = c \; , \qqquad 
  c_F = \frac {g_V^2}{2g^2} \, c \; , \qqquad 
  c_{HHVV} = c^2 \; .
\end{align}
This reduces the list of generated dimension-6 operators to
\begin{align}
  f_{WW} = f_{BW} = \frac {c^2} {2g^2} \qquad \text{and} \qquad  f_W = - \frac {c^2} {g^2} \,,
\end{align}
and all dimension-6 deviations scale like $c^2/m_\xi^2$. To avoid
effects from strongly interacting theories we limit our analysis to
$\Gamma_{\xi}/m_{\xi} < 1/4$.

In the left panel of Fig.~\ref{fig:limits} we see that based on event
numbers in the range $150~\text{GeV} < p_{T,j_1} < 300~\text{GeV}$,
the dimension-6 approximation with the squared terms gives the
same limits as the full model, as long as we ensure that the new
resonance remains narrow.  In the high-energy tail
$p_{T,j_1} > 300$~GeV including the squared terms also improves the validity of
the dimension-6 approach, but it only leads to identical limits for
large $m_\xi$, combined with strong couplings. Indeed, limiting the
momentum transfer of events for example through an upper limit on
$p_{T,j}$ is well known to reduce the dependence on model
assumptions~\cite{spins1,spins2}.

Just as for the $Vh$ production process, at least as long as the event
numbers remain small the square of the dimension-6 operators always
improves the agreement with the full theory in weak boson fusion. With
improved statistics the differences become smaller and ultimately
negligible, and the question of whether the squared dimension-6
amplitudes should be taken into account is rendered irrelevant.

\subsection*{Realistic tagging jets}

\begin{figure}[t]
  \includegraphics[width=0.43\textwidth]{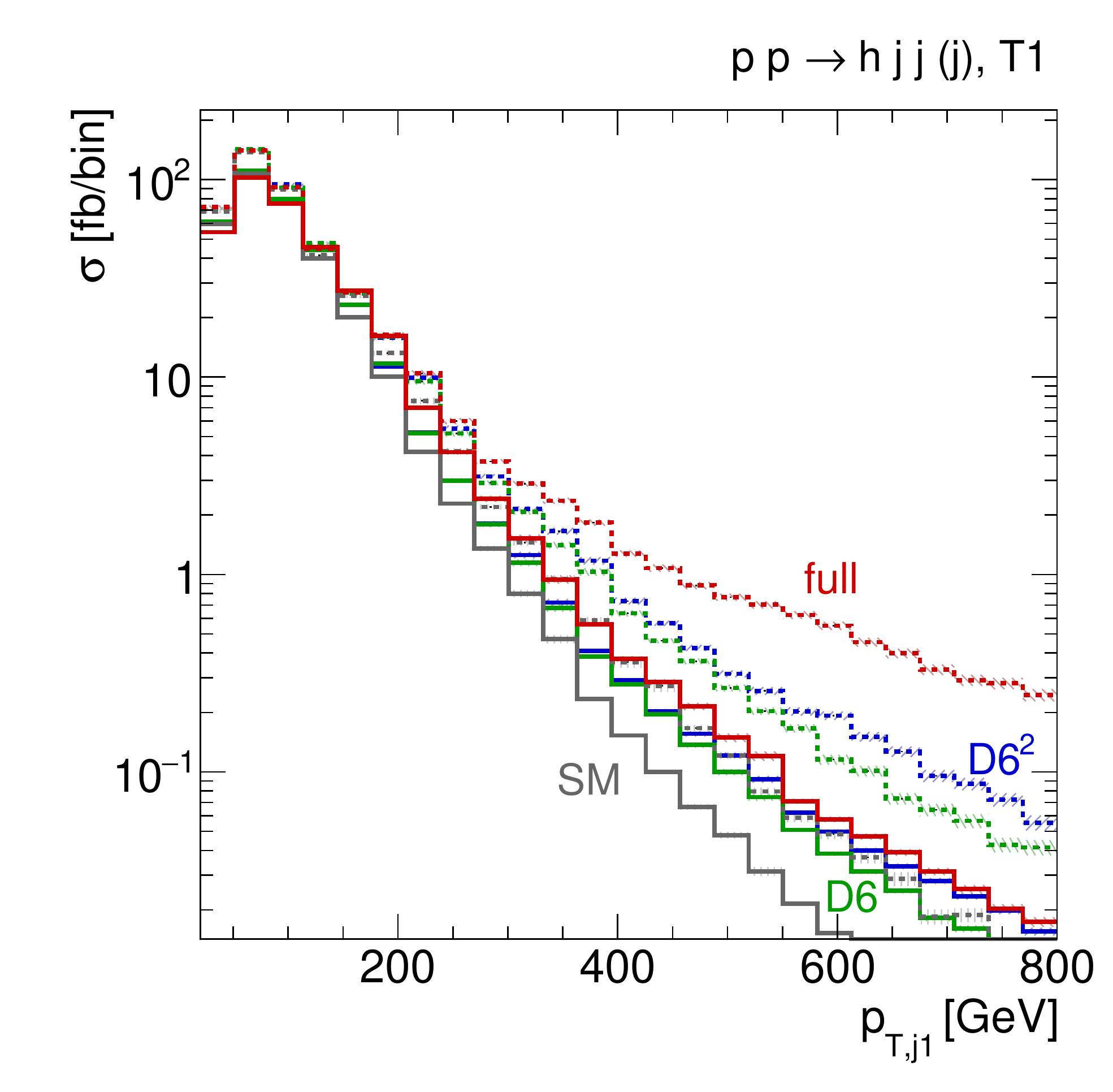}
  \hspace*{0.05\textwidth}
  \includegraphics[width=0.43\textwidth]{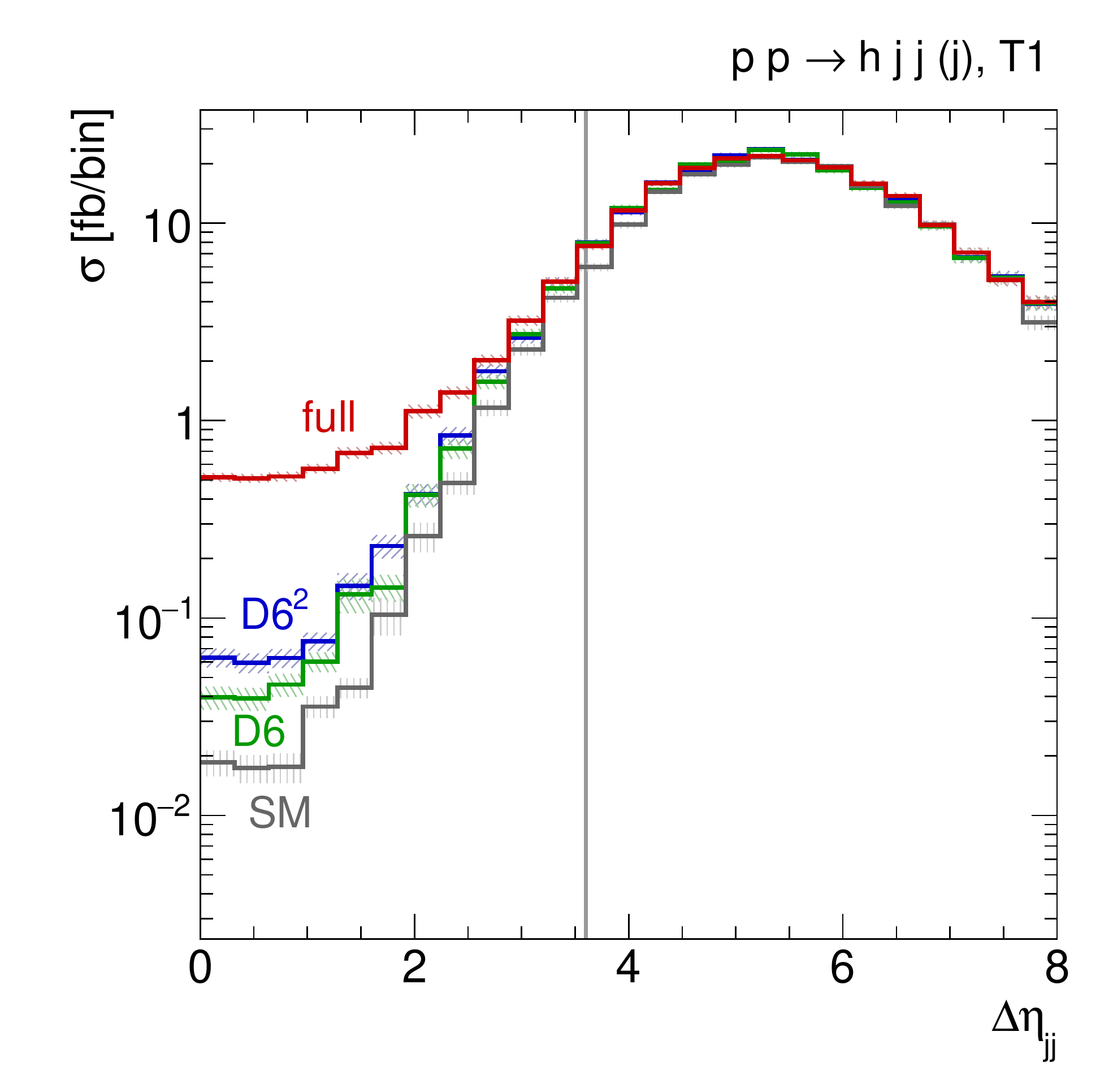}
  \caption{WBF distribution at hadron level. Left: $p_{T,j_1}$
    distribution based on the full process, the dashed lines show the
    distributions based on WBF diagrams only and without a
    $\Delta \eta_{jj}$ cut. Right: $\Delta \eta_{jj}$ based on WBF
    diagrams only, the vertical line marks the standard WBF cut
    following Eq.\;\eqref{eq:wbf_cuts}.}
  \label{fig:realistic_jets}
\end{figure}

Before we attempt to further improve the description of the full vector
triplet model for example in the benchmark point T1, we briefly test
if the parton-level effects described above survive a realistic environment.
We add a parton shower and jet reconstruction now for the full
process
\begin{align}
  p p \to h \; j j \, (+j) \; ,
\end{align}
simulated in \toolfont{MadGraph}~\cite{madgraph}.  Parton showering is
performed by \toolfont{PYTHIA6}~\cite{pythia} using the $k_T$-jet MLM
matching scheme~\cite{mlm} with a minimum $k_T$ jet measure between
partons of \toolfont{xqcut}=20~GeV. \toolfont{Fastjet}~\cite{fastjet}
is used to construct jets based on the $k_T$ algorithm with $R = 0.4$. We do not
include a Higgs decay because we are only interested in
production-side kinematics.  The standard WBF cuts then are
\begin{align}
  p_{T,j} > 20~\gev \,, \qqquad 
  m_{jj} > 500~\gev \,, \qqquad 
  \Delta \eta_{jj}~>~3.6
\label{eq:wbf_cuts}
\end{align}
on the two hardest jets. We veto additional jets with
$p_{T,j} > 20$~GeV between these two tagging jets.  To analyze the
effects of the $\Delta \eta_{jj}$ cut~\cite{spins2}, we generate
additional samples explicitly excluding Higgs-strahlung diagrams, in
spite of the fact that it might break gauge invariance.

In Fig.~\ref{fig:realistic_jets} we show that the distributions are
generally robust under parton shower and jet reconstruction, but two
complications arise.  First, on-shell $\xi$ production contributes to
this process and is not entirely removed by the WBF cuts in
Eq.\;\eqref{eq:wbf_cuts}, leading to visible differences between the
full and effective model already at low momenta. Such a resonance peak
would be easy to identify experimentally and does not present a major
problem for the dimension-6 approximation.

Second, the tension between the full model and the dimension-6
approximation at large momenta now remains below $10~\%$.  This means
that the $\Delta \eta_{jj}$ cut not only removes large contributions
from Higgs-strahlung--like diagrams, it also gets rid of phase-space
regions where the full model and the dimension-6 description differ
the most.  At the same time, the $\Delta \eta_{jj}$ removes some of
its well-known discrimination power for new physics effects versus the
Standard Model~\cite{spins2}.

\section{Towards a simplified model}
\label{sec:simplified}

\begin{figure}[t]
  \includegraphics[width=0.43\textwidth]{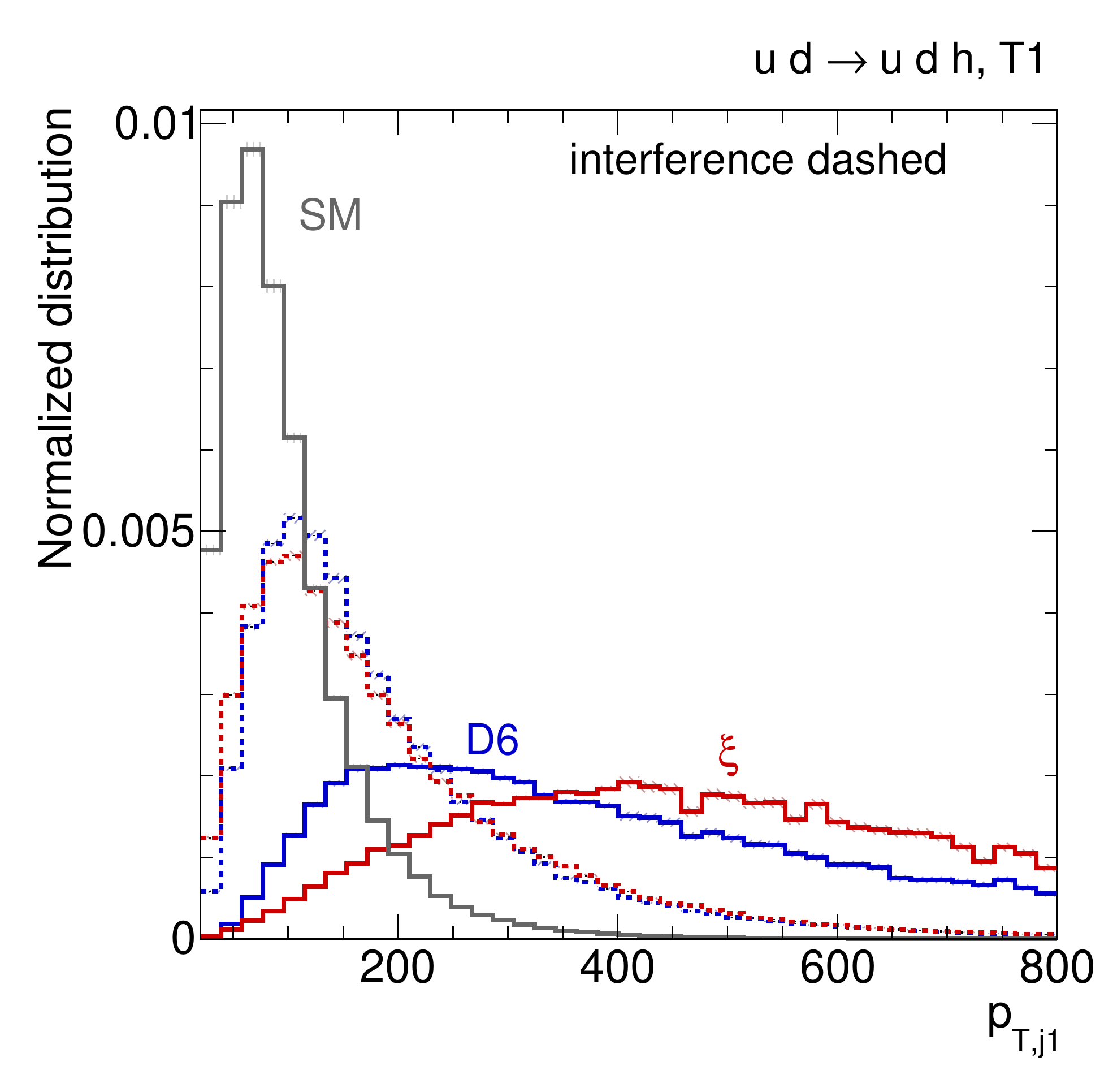} 
  \hspace*{0.05\textwidth}
  \includegraphics[width=0.43\textwidth]{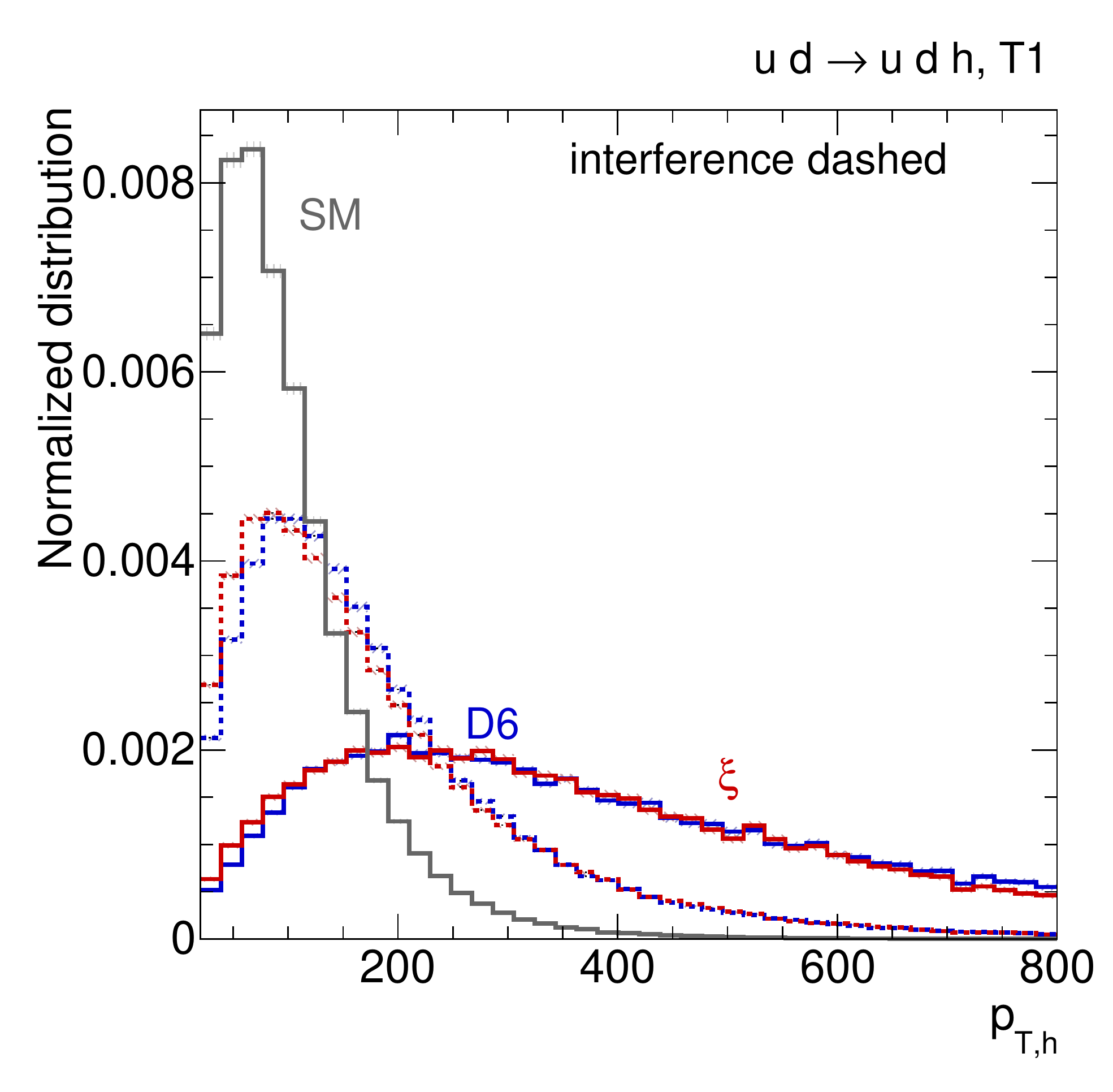}\\
  \includegraphics[width=0.43\textwidth]{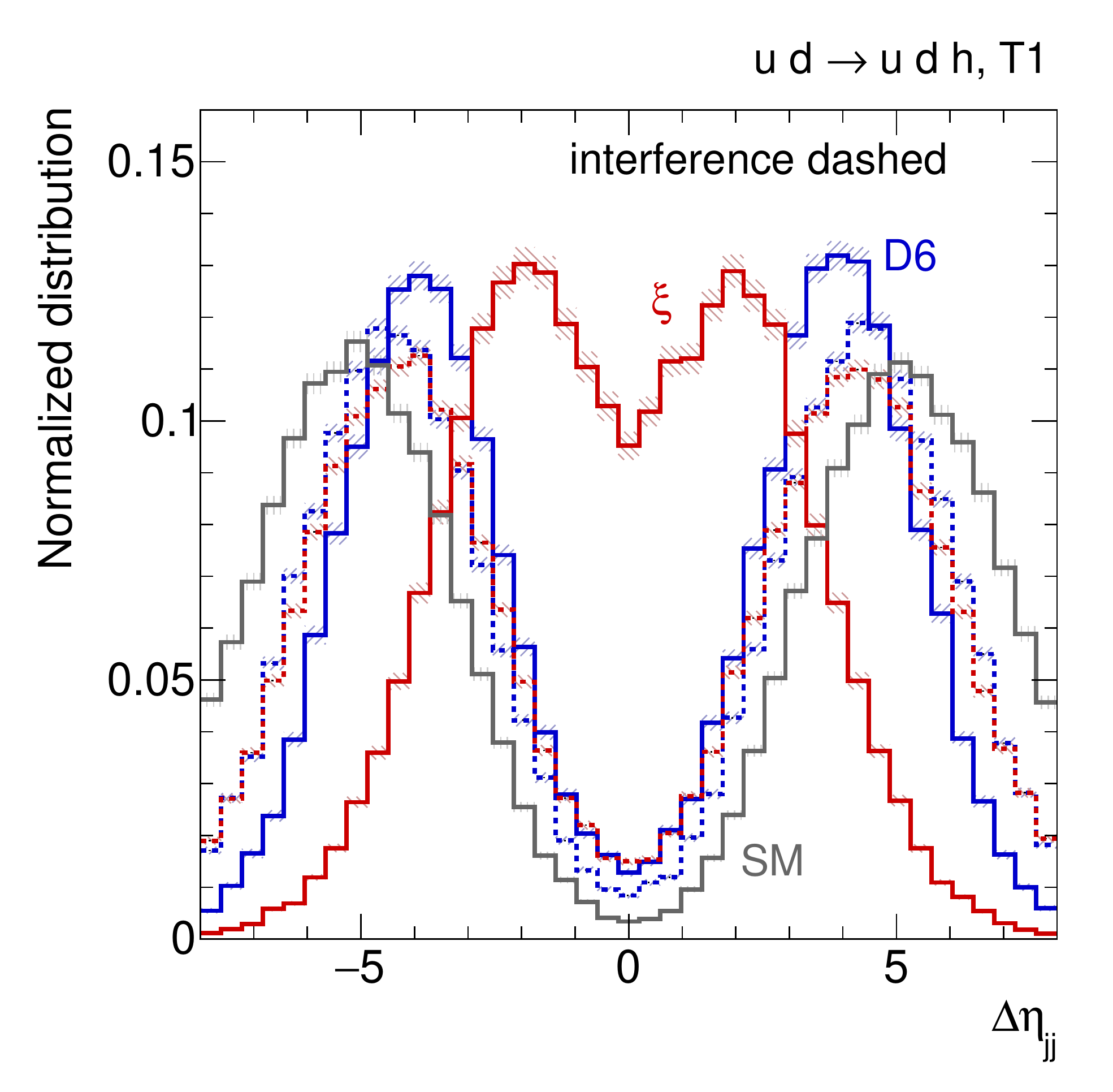} 
  \hspace*{0.05\textwidth}
  \includegraphics[width=0.43\textwidth]{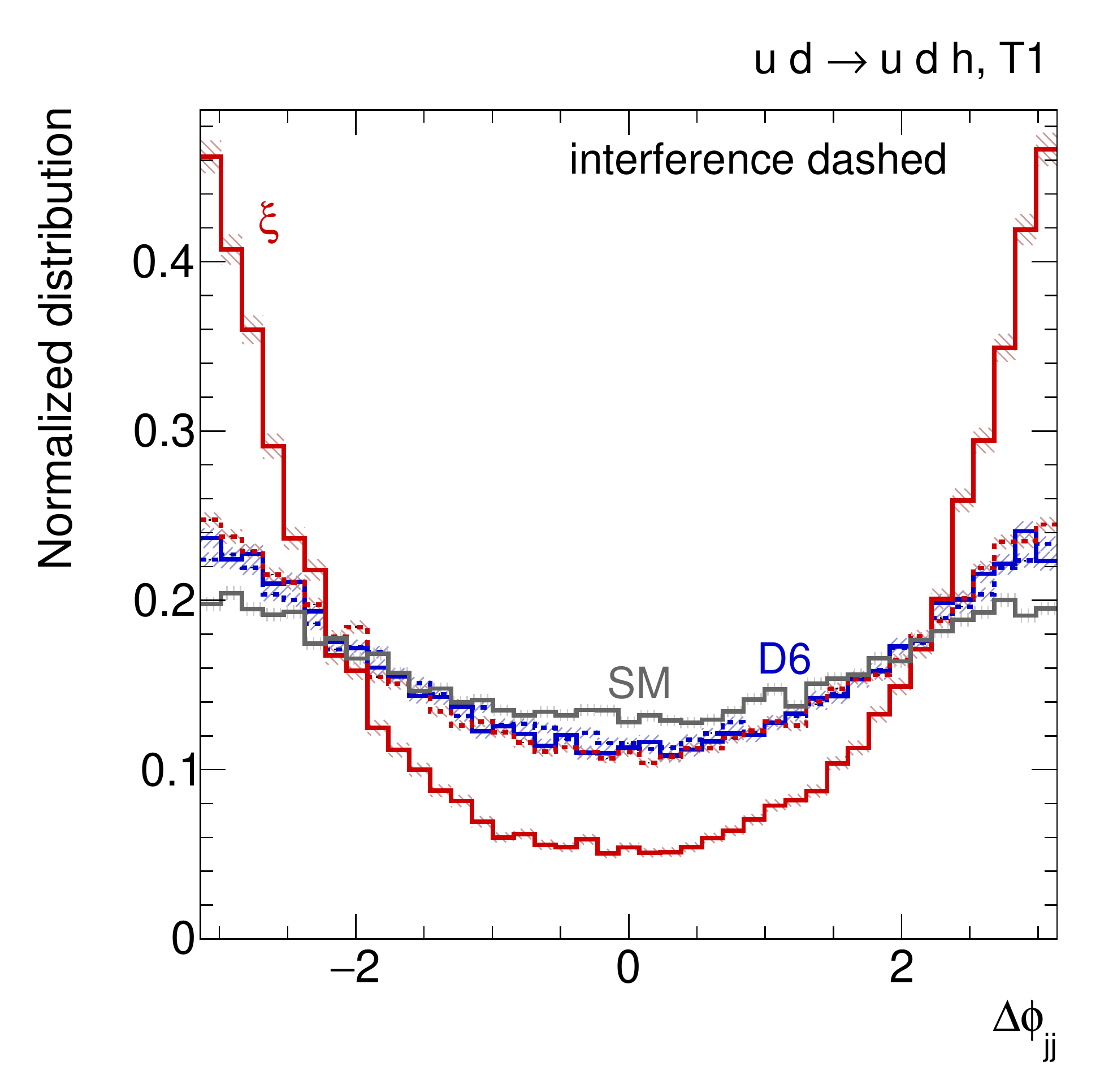}
  \caption{Normalized WBF distributions of the tagging jets. We separate
    the squared new-physics amplitudes, shown as solid lines, from the
    interference with the SM-like diagrams (dashed).}
  \label{fig:squared_separate}
\end{figure}

In the first part of the paper we have shown where in phase space a
dimension-6 description of LHC observables breaks down, both for $Vh$
production and for weak boson fusion. For $Vh$ production with its
simple $2 \to 2$ kinematics problems are clearly linked to a possible
$s$-channel resonance, as seen in Eq.\;\eqref{eq:breakdown_vh}.  For
weak boson fusion there appears no resonance, but the result of
Eq.\;\eqref{eq:breakdown_wbf} suggests that the new states in the
$t$-channel have a similar effect.  In Fig.~\ref{fig:squared_separate}
we show different tagging jet distributions, separating the Feynman
diagrams including the heavy $\xi$ states. In particular for the
critical $p_{T,j_1}$ distribution, the $\Delta \eta_{jj}$
distribution, and the $\Delta \phi_{jj}$ distribution these diagrams
are only very poorly described by the dimension-6 approach. In
practice this is not a problem because these contributions are
strongly suppressed by the heavy mass $m_\xi$, but it poses the
question how we can improve the agreement. The obvious solution to
these problems in the $s$-channel of $Vh$ production and in the
$t$-channel of weak boson fusion is a simplified
model~\cite{simp,simp_higgs}. A new vector field mixing with the weak
bosons as described by the Lagrangian shown in
Eq.\;\eqref{eq:lag-vectortriplet} is such a simplified model, but its
structure is still relatively complex. Obviously, an additional heavy
scalar with mass around $m_\xi$ and the appropriate couplings will
improve the $2 \to 2$ kinematics for $Vh$ production. The question we
want to study in this section is if such a scalar can also improve the
weak boson fusion kinematics.

\subsection*{A pseudo-scalar as a simplified vector}

The simplest simplified model we can write down includes one new
massive scalar $S$ with a Higgs portal and a Yukawa coupling. 
However, a scalar state will not interfere with the Standard Model
diagrams. In analogy to the CP properties of the Goldstone mode
contributing to the massive $Z$ boson we define our simplified model
with a pseudo-scalar state as
\begin{align}
\mathcal{L} \supset 
  \frac{1}{2} (\partial_\mu S)^2 
- \frac{m_S}{2} S^2 
+ \sum_\text{fermions} g_F \; S \overline{F} \gamma_5 F 
+ g_S \; S^2 \phi^\dagger \phi \,.
  \label{eq:simplified_model}
\end{align}

\begin{figure}[t]
  \includegraphics[width=0.43\textwidth]{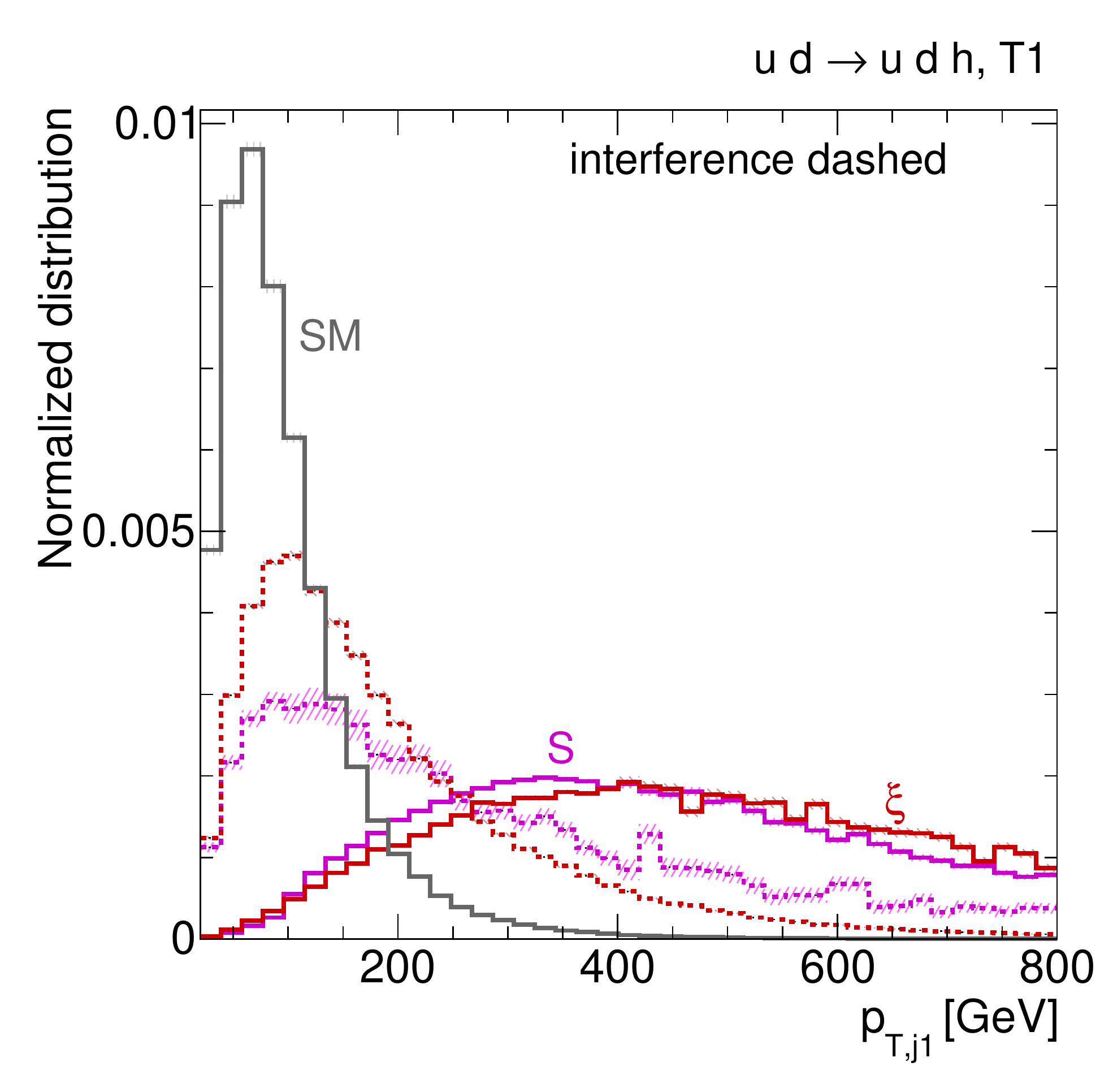}
  \hspace*{0.05\textwidth}
  \includegraphics[width=0.43\textwidth]{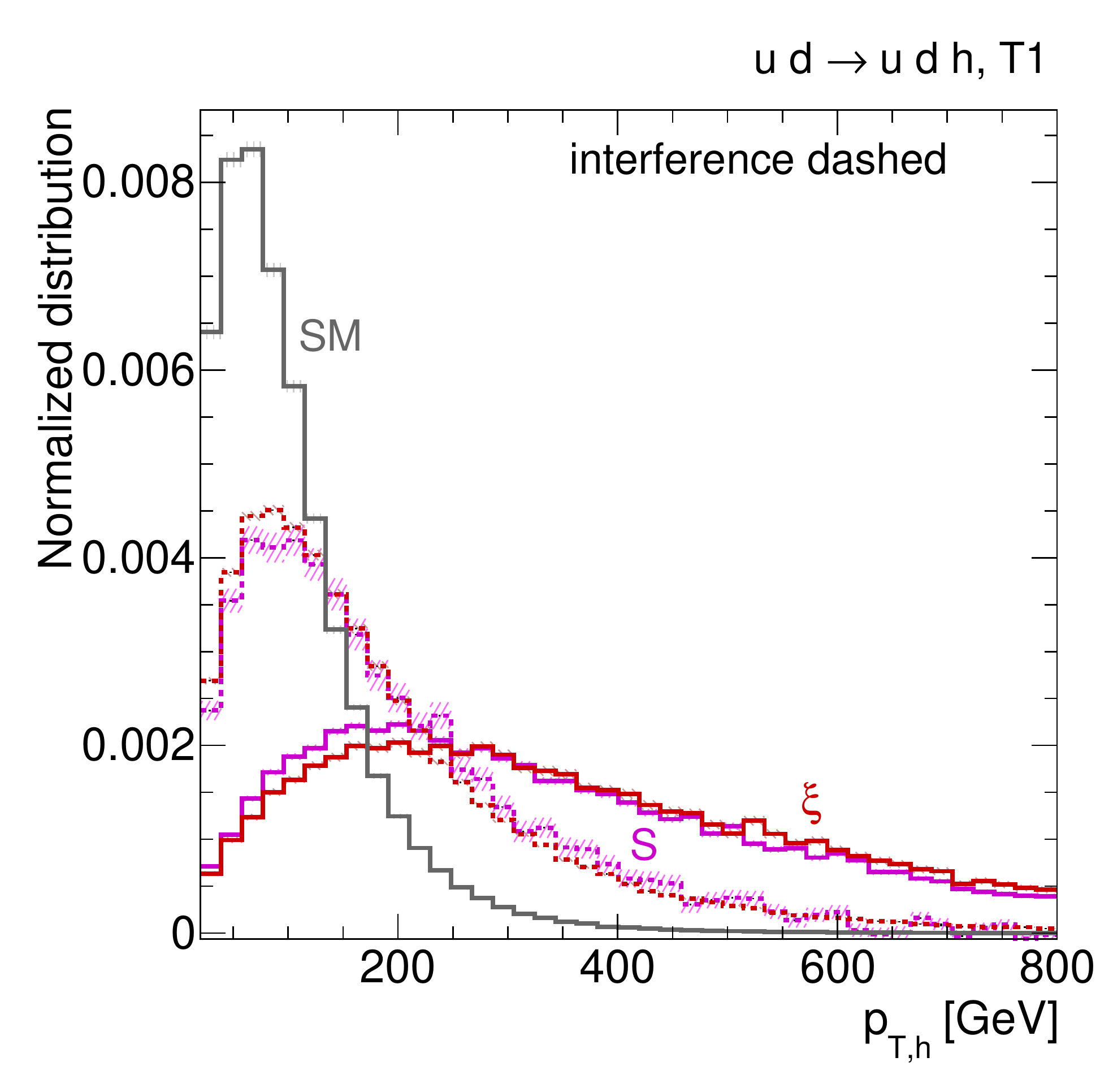} \\
  \includegraphics[width=0.43\textwidth]{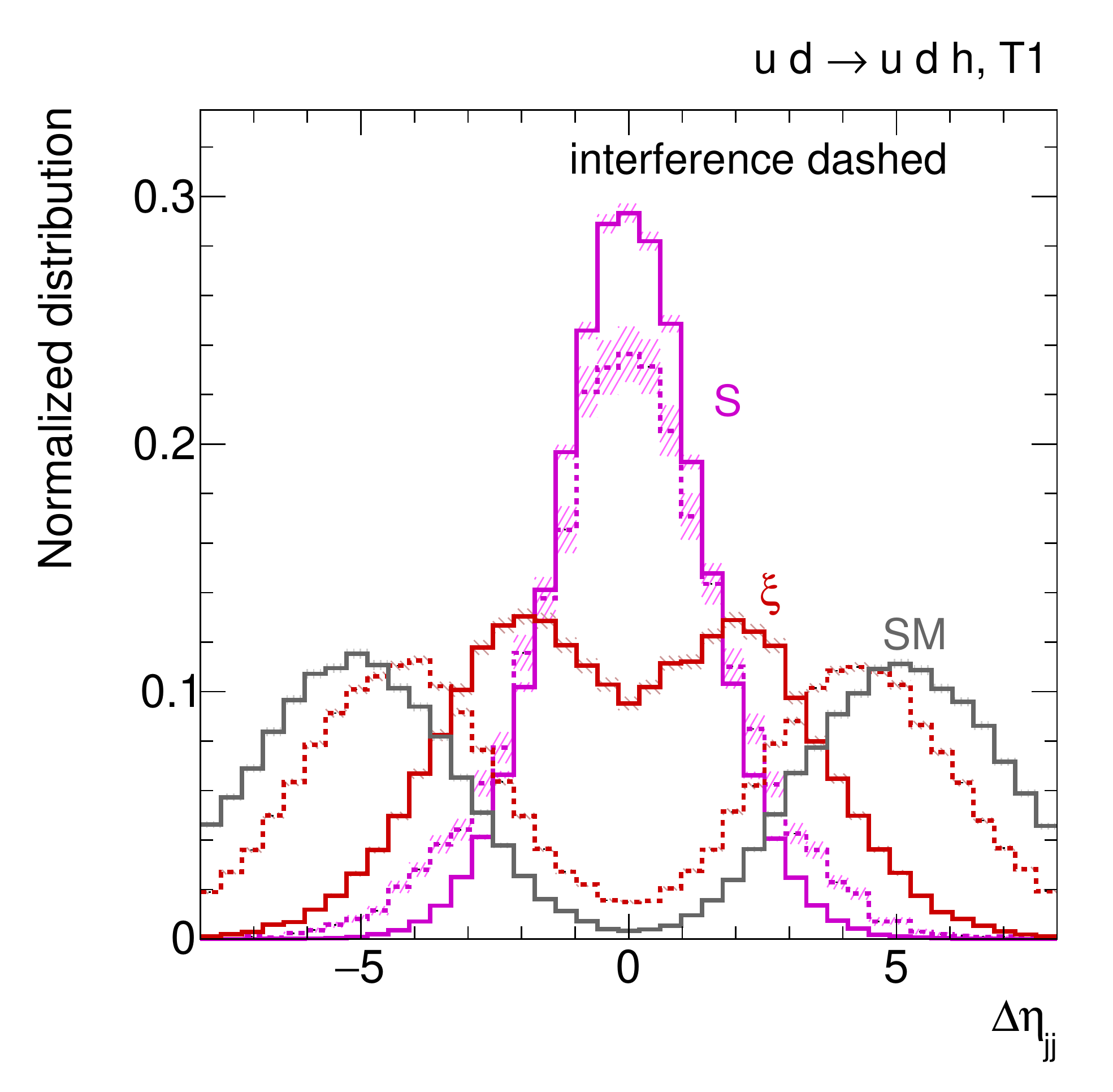}
  \hspace*{0.05\textwidth}
  \includegraphics[width=0.43\textwidth]{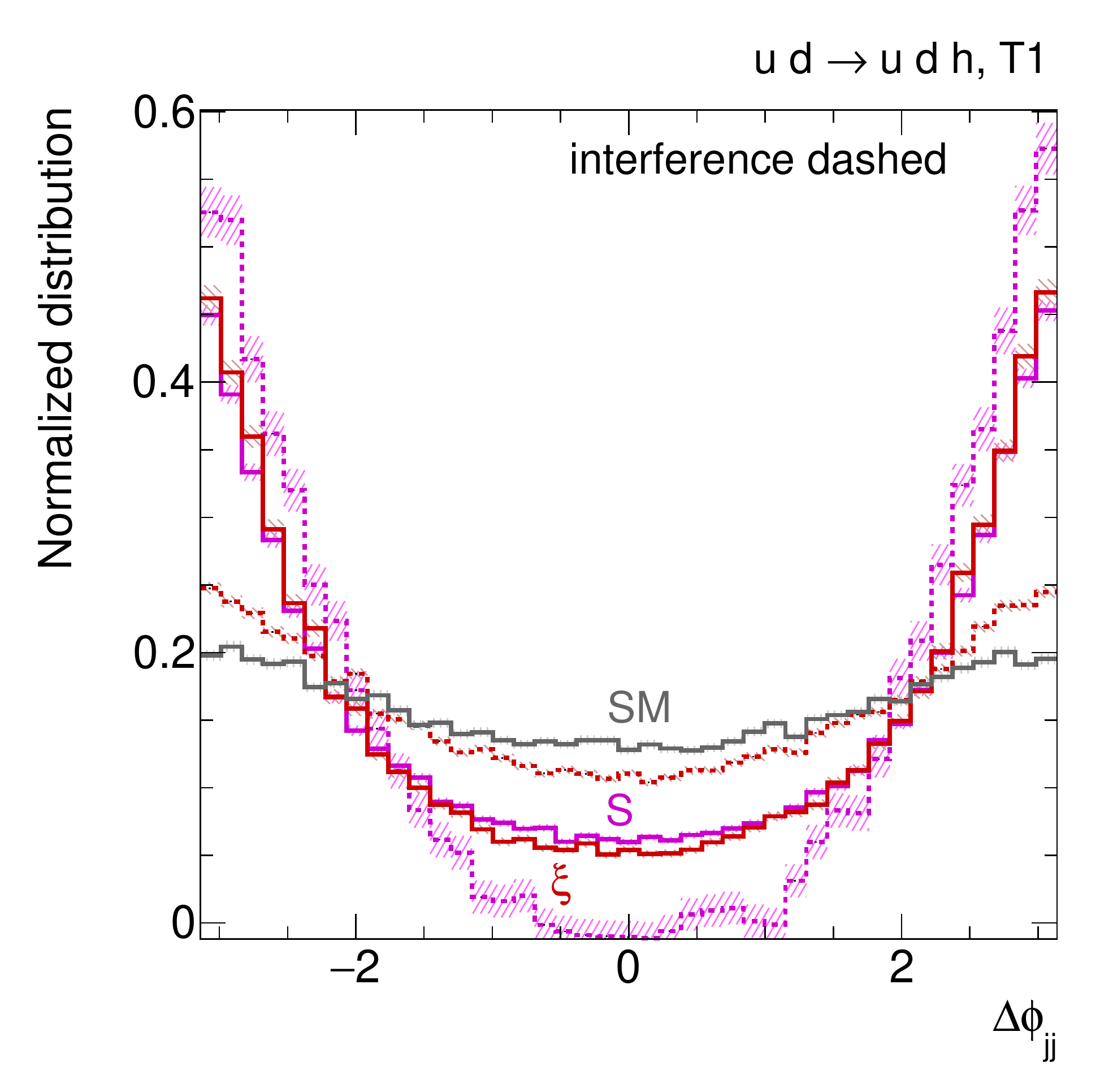}
  \caption{Normalized WBF distributions for a scalar simplified model
    defined in Eq.\;\eqref{eq:simplified_model} vs the vector triplet
    benchmark.}
  \label{fig:simplified}
\end{figure}

In Fig.~\ref{fig:simplified} we show the same WBF distributions as in
Fig.~\ref{fig:squared_separate}, but including the simplified scalar
model. For the $p_{T,j}$ distribution the squared new-physics
amplitudes in the full vector model and the simplified scalar model
indeed agree well, improving upon the dimension-6 description which
breaks down in this distribution.  However, the interference term with
the Standard Model, which is numerically dominant for most of the
distribution and well described in the dimension-6 model, poses a
problem.  The $\Delta \eta_{jj}$ distributions show even poorer
agreement: the spin-1 amplitudes of the Standard Model and the vector
triplet have similar phase-space distributions and give two forward
tagging jets, while the scalar mediator favors central
jets~\cite{spins2}.  The $\Delta \phi_{jj}$ distribution, known to be
sensitive to the tensor structure of the hard $VVh$
interaction~\cite{delta_phi}, exposes similar differences between the
full and simplified model.  Altogether, our simplified scalar model
with its very different $VVh$ interaction structure does improve the
description in the region where the dimension-6 approach breaks down,
but it fails to describe interference patterns and angular
correlations of the tagging jets.

\subsection*{Splitting functions and equivalence theorem}

We can understand this very different behavior of the scalar
$t$-channel mediator as compared to the vector from the splitting
kernels in the collinear limit.  The matrix element squared for the
weak boson fusion process mediated by pseudo-scalars $S$ has the form
\begin{align}
 | \mathcal{M}(qq \to q'q'h) |^2 \propto 
  \frac{g_F^4 \;  t_1 t_2}{(t_1 - m_S^2)^2 \; (t_2 - m_S^2)^2} 
\stackrel{m_S \to 0}{\longrightarrow} \frac{\text{const}}{t_1 t_2} \; ,
\end{align}
where $t_1$ and $t_2$ denote the respective momentum flow through each
scalar propagator. For $m_S \to 0$ the Jacobians from the phase-space
integration cancel a possible collinear divergence, while for a light
vector boson a soft and a collinear divergence remains. Unlike in the
usual WBF process, the tagging jets in our simplified scalar model
will not be forward.  The reason for this difference in the
infrared is the (pseudo-)scalar coupling to quarks: since the scalar
carries no Lorentz index, a $q \to q S$ splitting will be expressed in
terms of the momentum combinations $(p_q p_q')$, $p_q^2 = m_q^2$, and
$p_q'^2 = m_q^2$. In the limit of massless quarks only the first term
remains as $t = 2 (p_q p_q')$.  This factor in the numerator cancels
the apparent divergence of the $t$-channel propagator.

Adding higher-dimensional couplings of the (pseudo-)scalar to
fermions, such as
\begin{align}
  \mathcal{L} \supset 
\sum_\text{fermions} \Biggl[  
  g_{F,2} S \overline{F} F 
+ g_{F,3} (\partial_\mu S) \overline{F} \gamma^\mu F
+ g_{F,4} S (\partial_\mu S) \overline{F} \gamma^\mu \gamma_5 F 
+ g_{F,5} S (\partial_\mu \partial_\nu S) \overline{F} [\gamma^\mu,\gamma^\nu] F
\Biggr] \; ,
\label{eq:simplified_model_extended}
\end{align}
does not change this result qualitatively. After partial integration
and using the Dirac equation for the on-shell quarks the coupling
$g_{F,3}$ is equivalent to the simple scalar coupling, $g_{F,2} = m_q^2
g_{F,3}$. In the limit of massless quarks, only two of the new
structures listed in Eq.\;\eqref{eq:simplified_model_extended}
contribute at all: $g_{F,2}$ gives exactly the same result as $g_F$,
while $g_{F,5}$ leads to even higher powers of $t$ in the numerator,
\begin{align}
  | \mathcal{M}(qq \to q'q'h) |^2 \propto 
  \frac{g_{F,5}^4 \; t_1^3 t_2^3}{(t_1 - m_s^2)^2 \; (t_2 - m_s^2)^2} \; . 
\end{align}
No matter how we couple the (pseudo-)scalar of the simplified model to
the external quarks, it never reproduces the collinear splitting
kernel of a vector boson.

To be a little more precise, we can write out the spin-averaged matrix
element squared for the $q \to q' S$ splitting in terms of the energy
of the initial quark $E$, the longitudinal momentum fraction $x$, and
the transverse momentum $p_T$, both carried by $S$,
\begin{align}
 | \mathcal{M}(q \to q'S) |^2 &= - 2 g_F^2 x m_q^2
                     + 2 g_F^2 E^2 (1-x)
                     \Biggl[ \sqrt{1 + \frac {p_T^2} {E^2 (1-x)^2} + \frac {m_q^2 (1 - (1-x)^2)} {E^2 (1-x)^2} } - 1 \Biggr] \notag \\
                   &= g_F^2 \, \frac {x^2 \, m_q^2} {1-x} 
                     + g_F^2 \,  \frac {p_T^2} {1-x} 
                     + \ord\left(\frac{m_q^2 p_T^2}{E^2}, \frac{m_q^4}{E^2}, \frac{p_T^4}{E^2}\right) \;.
\label{eq:splitting_s}
\end{align}
From Eq.\;\eqref{eq:splitting_s} one can derive an effective Higgs
approximation or \emph{effective scalar
  approximation}~\cite{effective_scalar}: in the collinear and
high-energy limit, a process $q X \to q' Y$ mediated by a
(pseudo-)scalar $S$ is described by
\begin{align}
  \sigma (qX \to q'Y) = \int \mathrm{d}x \, \mathrm{d} p_T \, F_S(x,p_T)
  \, \sigma (SX \to Y)
\label{eq:def_splitting}
\end{align}
with the splitting function
\begin{align}
  F_S(x,p_T) &= \frac {g_F^2} {16 \pi^2} \, 
               \frac {x \, p_T^3} {\left( m_S^2 (1-x) + p_T^2 \right)^2} \,.
\label{eq:kernel_s}
\end{align}
Unlike for vector emission, there is no soft divergence for $x \to 0$.
The $p_T$ dependence is the same as for transverse vector
bosons~\cite{effective_w,polarized_ww}, as we discuss in some detail in the
appendix. 

It might seem surprising that our pseudo-scalar is emitted with a
fundamentally different phase-space dependence than longitudinal $W$
and $Z$ bosons, in apparent contradiction of the Goldstone boson
equivalence theorem.  However, the latter only makes a statement about
the leading term in an expansion in $m_W / E$, where 
$\varepsilon^\mu_L \sim p^\mu / m_W$. At this order the squared matrix
element for the splitting $q \to q' W_L$ agrees with the pseudo-scalar
result, but is suppressed by a factor of $m_q^2 / E^2$. Higher orders
in the $m_W/E$ expansion, outside the validity range of the
equivalence theorem, are not suppressed by quark masses.  The
equivalence theorem is therefore of very limited use in describing the
$W$ or $Z$ couplings to quarks except the top.

\subsection*{Which observable to study}

\begin{figure}[t]
  \includegraphics[width=0.43\textwidth]{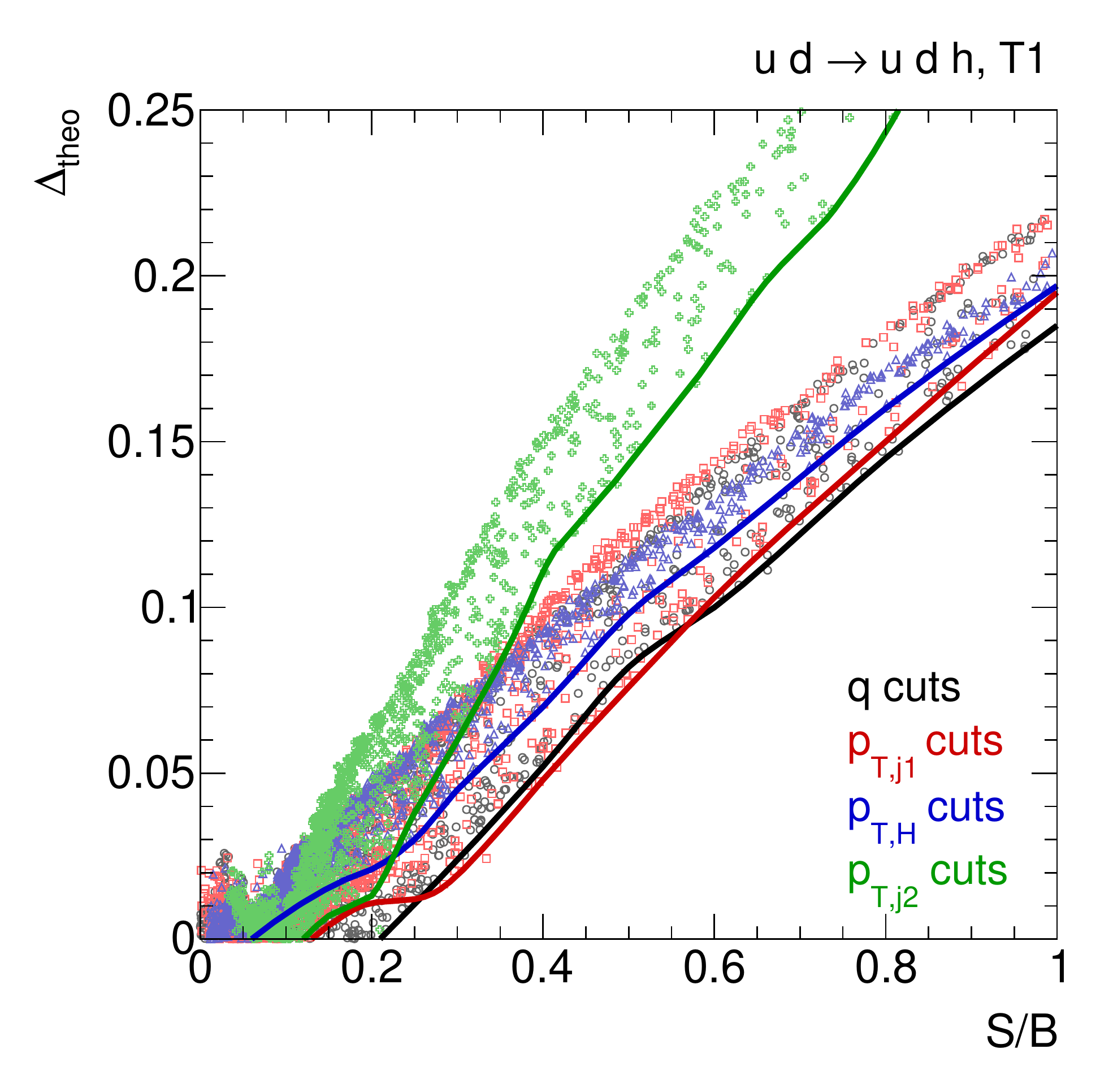}
  \hspace*{0.05\textwidth}
  \includegraphics[width=0.43\textwidth]{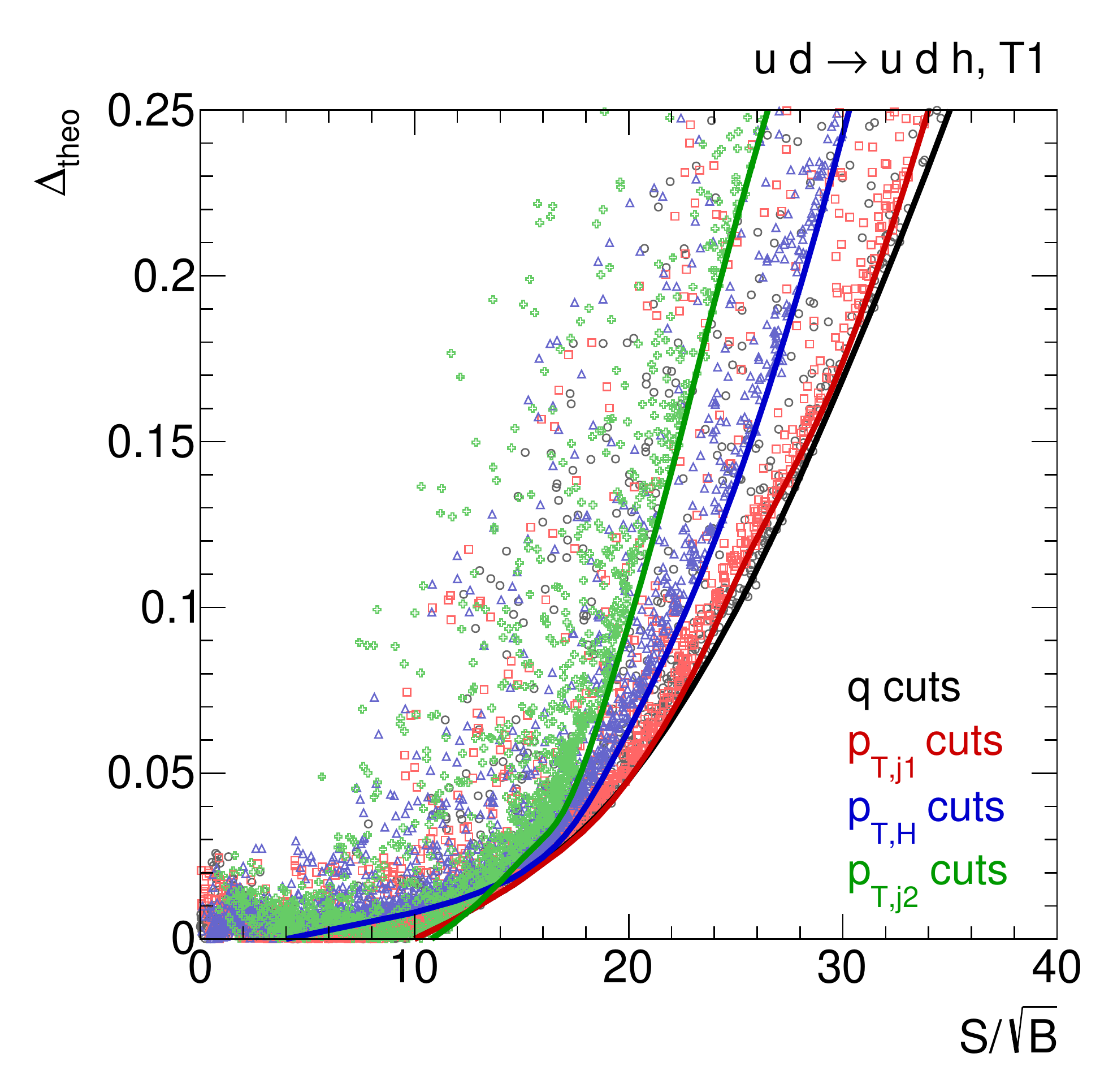}
  \caption{Experimental reach in systematics-driven and
    statistics-driven channels vs theoretical uncertainties of the
    dimension-6 description. Each point corresponds to a window
    $x_\text{min,max}$ in one of the four momentum observables that leaves a signal
    cross section of at least 20~fb.}
  \label{fig:cuts}
\end{figure}

Now that it is clear that we cannot further improve the agreement
between the vector triplet and its dimension-6 approximation by adding
a heavy scalar as a simplified model, we go back to the original
problem: how can we best use the dimension-6 approximation for limit
setting, and do the shortcomings shown in
Fig.~\ref{fig:squared_separate} harm this approach?

We know that in our LHC analysis we should avoid angular correlations
of the tagging jets, like $\Delta \eta_{jj}$ or $\Delta
\phi_{jj}$. Instead, we can use momentum-related kinematic variables
like
\begin{align}
 x \in \left\{  q, \, p_{T,j_1}, \, p_{T,j_2}, \, p_{T,h} \right\} \; .
\label{eq:variables}
\end{align}
An acceptance cut $x > x_\text{min}$ on any of those variables
projects out the interesting phase-space regions, while the cut $x <
x_\text{max}$ ensures the validity of an effective theory
description. If $x_\text{min} > x_\text{max}$ the dimension-6
description is not useful. For each window $x_\text{min,max}$ we can
compute the contribution to the theoretical uncertainty
\begin{align}
  \Delta_\text{theo} (x_\text{min,max}) 
= \left| \frac {\sigma_\text{D6} - \sigma_\text{full}} {\sigma_\text{full}} \right| \; ,
\label{eq:err_th}
\end{align} 
as well as the statistics-driven and systematics-driven significances
\begin{align}
  \frac{S}{B} (x_\text{min,max}) 
= \left| \frac {\sigma_\text{full} - \sigma_\text{SM}} {\sigma_\text{SM}} \right| 
\qquad \text{and} \qquad 
  \frac{S}{\sqrt{B}} (x_\text{min,max}) 
= \sqrt{L} \, \left| \frac {\sigma_\text{full} - \sigma_\text{SM}} {\sqrt{\sigma_\text{SM}}} \right| \; ,
\label{eq:err_ex}
\end{align}
where $L = 30~\ifb$ is used as a toy number.

The question is for which
observable $x$ we find the largest $S/B$ and $S/\sqrt{B}$ values while
keeping $\Delta_\text{theo}$ small.  In Fig.~\ref{fig:cuts} we show
the correlations between theoretical uncertainty and experimental
reach for the variables defined in Eq.\;\eqref{eq:variables} for a
parton-level analysis as defined in Eq.\;\eqref{eq:def_wbf}. We see
that the momentum transfer $q$ or the leading tagging jet's
$p_{T,j_1}$ lead to the envelopes with the highest significance for a
given theoretical uncertainty $\Delta_\text{theo}$. This indicates
that the leading tagging jet's transverse momentum is the best way of
experimentally accessing the momentum flow through the hard process,
at least for the hard parton-level process with only two tagging
jets.

\section{Summary}
\label{sec:summary}

While a dimension-6 Higgs analysis at the LHC cannot be considered the
leading part of a consistent effective theory, it describes the
effects of weakly interacting extensions of the Higgs-gauge sector very
well~\cite{too_long}. In this brief study we have answered two practical
question concerning such a dimension-6 analysis for Run~II.

First, a priori it is not clear if squared dimension-6 terms should be
included in calculations. We have studied two particularly challenging
parameter points of a vector triplet model for $Vh$ production and for
weak-boson-fusion Higgs production. For both processes we find that
the dimension-6 squared term avoids negative rate predictions in the
$m_{Vh}$ or $p_{T,V}$ distributions of $Vh$ production and in the
$p_{T,j_1}$ distribution of weak boson fusion. Even for cases with a
constructive interference between the dimension-6 and the Standard
Model contributions, it turns out that including the dimension-6
squared term improves the agreement of kinematic distributions between
the full model and the dimension-6 approximation. Ultimately, this
translates into a better agreement in the expected exclusion
limits, and similar conclusions in a different framework have recently
been published in Ref.~\cite{gino}.

Second, we have attempted to improve the agreement between the full
model and our approximation by using a simplified model. The only
significantly simpler model than a mixing gauge extension is an
extended scalar sector. While the corresponding deviations between the
full model and the dimension-6 approximation are phenomenologically
hardly relevant, we find that such an additional scalar improves the
modelling of kinematic distributions of the kind $m_{Vh}$ and
$p_{T,j_1}$ where the dimension-6 description breaks down. However,
this comes at the cost of significant deviations in the dominant
interference terms. Moreover, once we include angular correlations
like $\Delta \eta_{jj}$ or $\Delta \phi_{jj}$ in weak boson fusion,
the simplified model fails badly. The difference can be traced to the
divergence structure of the corresponding splittings.

Seeing that the dimension-6 approach is still the better simple model
to describe new physics in WBF distributions, we have finally analyzed
which phase-space regions provide an interesting window to new physics
while being well described by the dimension-6 approximation.  We have
demonstrated that the leading tagging jet's $p_T$ distribution is
particularly suited for such a search for new physics.

\subsubsection*{Acknowledgments}

We would like to thank Torben Schell for very useful discussions and
Michael Kr\"amer for his encouragement and support. Moreover, we are
grateful to David Lopez-Val and Ayres Freitas, whose inexhaustible
energy prepared the ground for this study.  All authors acknowledge
support from the German Research Foundation (DFG) through the
Forschergruppe `New Physics at the Large Hadron Collider' (FOR 2239),
J.\,B.\ also through the Graduiertenkolleg `Particle physics beyond
the Standard Model' (GRK 1940).

\appendix
\section{Effective scalar approximation}

In Sec.~\ref{sec:simplified} we have introduced a pseudo-scalar in the
$t$-channel of weak boson fusion to describe some of the features
which we find in the full vector triplet model and which our
dimension-6 description does not describe well. In this appendix we
collect some of the main formulas and compare the kinematics of
fermions radiating scalars, transverse, or longitudinal gauge
bosons. Our formalism follows the effective
$W$ approximation~\cite{effective_w} as well as the effective Higgs
approximation~\cite{effective_scalar} and allows us to analytically
describe the soft and collinear behavior. If we do not need to
describe interference terms with SM gauge bosons we can start with a
CP-even scalar splitting $q \to qS$, in terms of the energy of the
initial quark $E$, the longitudinal momentum fraction $x$, carried by $S$, and the
scalar's transverse momentum $p_T$:
\begin{align}
 | \mathcal{M}(q \to q'S)  |^2 &= 2 g_F^2 (2-x) m_q^2
                     + 2 g_F^2 E^2 (1-x)
                     \Biggl[ \sqrt{1 + \frac {p_T^2} {E^2 (1-x)^2} + \frac {m_q^2 (1 - (1-x)^2)} {E^2 (1-x)^2} } 
                       - 1 \Biggr] \notag \\
                   &= g_F^2 \left( 4  + \frac {x^2} {1-x} \right) m_q^2
                     + g_F^2 \, \frac {p_T^2} {1-x} 
                     + \ord\left(\frac{m_q^2 p_T^2}{E^2}, \frac{m_q^4}{E^2}, \frac{p_T^4}{E^2}\right) \; .
\end{align}
The main feature of this splitting is that the infrared behavior is
different for the term proportional to the quark mass and for the
surviving term in the realistic limit $m_q \to 0$: in the absence of a
fermion mass the collinear divergence from a $t$-channel propagator is
cancelled by the coupling structure. If the term proportional to $m_q$
dominates there will be the usual collinear divergence once we include
a scalar propagator. For a pseudo-scalar the structure shown in
Eq.\;\eqref{eq:splitting_s} is very similar,
\begin{align}
 |\mathcal{M}(q \to q'S)  |^2 &= - 2 g_F^2 x m_q^2
                     + 2 g_F^2 E^2 (1-x)
                     \Biggl[ \sqrt{1 + \frac {p_T^2} {E^2 (1-x)^2} + \frac {m_q^2 (1 - (1-x)^2)} {E^2 (1-x)^2} } 
                                - 1 \Biggr] \notag \\
                   &= g_F^2 \, \frac {x^2 \, m_q^2} {1-x} 
                     + g_F^2 \,  \frac {p_T^2} {1-x} 
                     + \ord\left(\frac{m_q^2 p_T^2}{E^2}, \frac{m_q^4}{E^2}, \frac{p_T^4}{E^2}\right) \;.
\end{align}

In the limit $m_q \to 0$ we can compute universal splitting kernels
including only the leading term in $p_T$, as defined in
Eq.\;\eqref{eq:def_splitting}.  Obviously, the scalar and pseudoscalar
case given in Eq.\;\eqref{eq:kernel_s} are identical, and we can compare
them with the splitting kernels for longitudinal or transverse
$W$ bosons~\cite{effective_w},
\begin{align}
  F_S(x,p_T) &= \frac {g_F^2} {16 \pi^2} \, x \,
               \frac {p_T^3} {\left( m_S^2 (1-x) + p_T^2 \right)^2} \,,\notag \\
  F_T(x,p_T) &= \frac {g^2} {16 \pi^2} \, \frac {1+(1-x)^2} x \, \frac {p_T^3} {\left( m_W^2 (1-x) + p_T^2 \right)^2} \,, \notag \\
  F_L(x,p_T) &= \frac {g^2} {16 \pi^2} \, \frac {(1-x)^2} x \, \frac {2 m_W^2 \, p_T} {\left( m_W^2 (1-x) + p_T^2 \right)^2} \,.
  \label{eq:splittings}
\end{align}

In Fig.~\ref{fig:effective_scalar} we show how these different
splittings translate into WBF distributions and compare full simulations
in \toolfont{MadGraph} to the predictions of Eq.\;\eqref{eq:splittings}.
A heavy Higgs, $m_h = 1$~TeV, is needed to guarantee a large energy scale
$E \sim m_h \gg p_T \sim m_W, m_S$. In this case we find that the
effective scalar approximation quite accurately describes the transverse
momentum distribution of the tagging jets. For $m_h = 125$~GeV the
assumption of on-shell $W$ bosons or scalars breaks down and the
effective descriptions lose their validity.

\begin{figure}[t]
  \centering
  \includegraphics[width=0.43\textwidth]{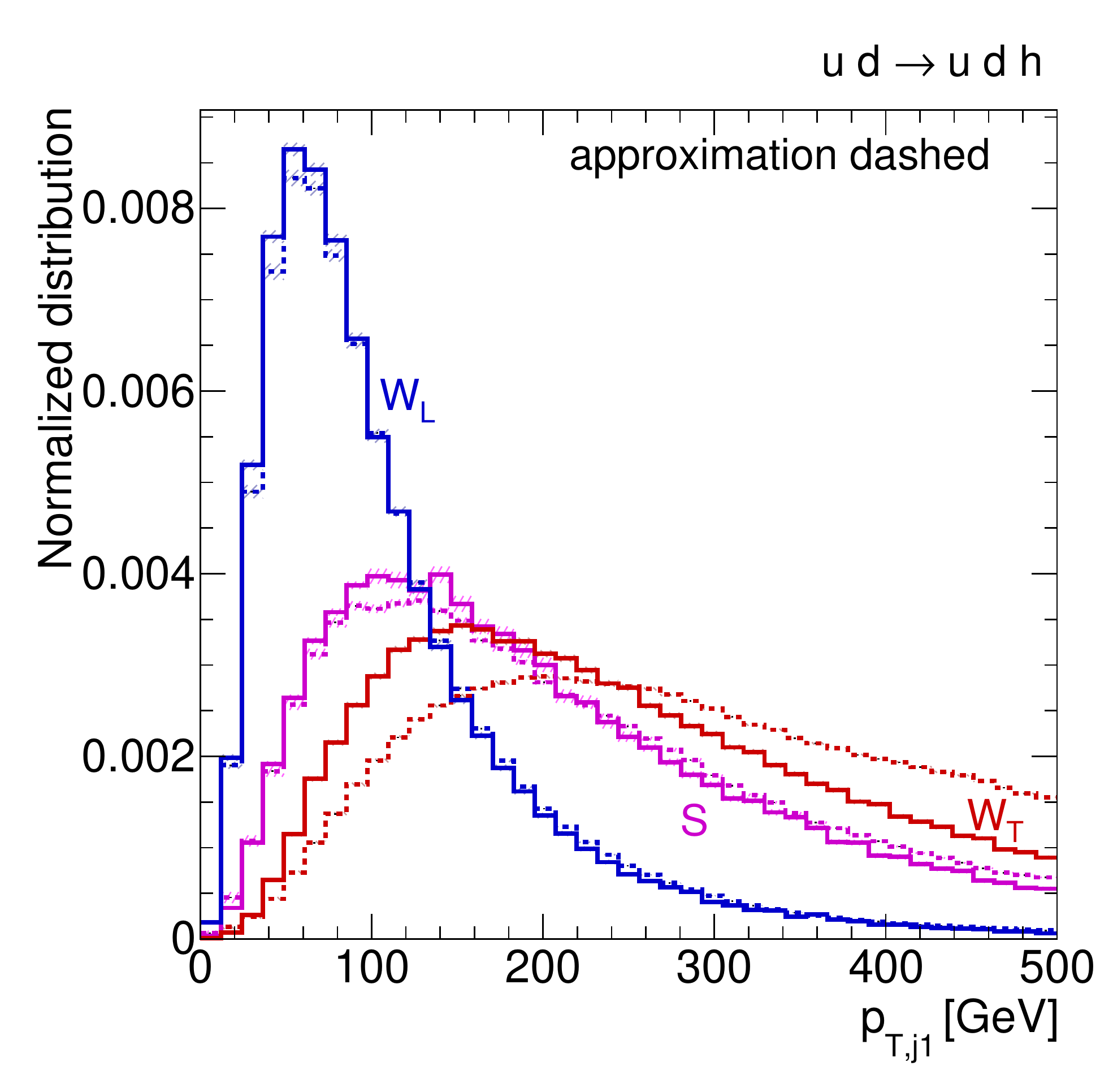}
  \caption{Normalized WBF distributions of the tagging jets in the SM with
  a heavy Higgs, $m_h = 1$~TeV. Scalar mediators are compared to
  longitudinal and transverse $W$ bosons following
  Ref.~\cite{polarized_ww}.
  The dotted lines give the corresponding predictions of the effective
  $W$ and scalar approximations, Eq.\;\eqref{eq:splittings}.}
  \label{fig:effective_scalar}
\end{figure}


\end{document}

%% file: declare.tex
\newcommand{\toolfont}[1]{\texttt{#1}}
\newcommand{\ord}{\ensuremath{\mathcal{O}}}
\newcommand{\ope}[1]{\ensuremath{\mathcal{O}_{#1}}}

\newcommand{\lag}{\ensuremath{\mathcal{L}}}
\newcommand{\mat}{\ensuremath{\mathcal{M}}}

\newcommand{\br}{\text{BR}}
\newcommand{\qqquad}{\qquad \qquad}

\newcommand{\gev}{{\ensuremath\rm GeV}}

\newcommand{\ifb}{{\ensuremath\rm fb^{-1}}}

\def\slashchar#1{\setbox0=\hbox{$#1$}           
   \dimen0=\wd0                                 
   \setbox1=\hbox{/} \dimen1=\wd1               
   \ifdim\dimen0>\dimen1                        
      \rlap{\hbox to \dimen0{\hfil/\hfil}}      
      #1                                        
   \else                                        
      \rlap{\hbox to \dimen1{\hfil$#1$\hfil}}   
      /                                         
   \fi}

\def\eg{{e.\,g.}\ }

\newcommand{\pbp}{\ensuremath{\phi^\dagger\,\phi}}

%% file: EFT_over_the_edge.bbl
\begin{thebibliography}{99}

\bibitem{higgs}
  P.~W.~Higgs,
  Phys.\ Lett.\  {\bf 12}, 132 (1964);
  P.~W.~Higgs,
  Phys.\ Rev.\ Lett.\  {\bf 13}, 508 (1964);
  F.~Englert and R.~Brout,
  Phys.\ Rev.\ Lett.\  {\bf 13}, 321 (1964).
  
\bibitem{discovery}
  G.~Aad {\it et al.} [ATLAS Collaboration],
  Phys.\ Lett.\ B {\bf 716}, 1 (2012);
  S.~Chatrchyan {\it et al.}  [CMS Collaboration],
  Phys.\ Lett.\ B {\bf 716}, 30 (2012).

\bibitem{bsmreview} 
  D.~E.~Morrissey, T.~Plehn and T.~M.~P.~Tait,
  Phys.\ Rept.\  {\bf 515}, 1 (2012).

\bibitem{eftfoundations}
  S.~Weinberg,
  Phys.\ Lett.\ B {\bf 91}, 51 (1980);
  S.~R.~Coleman, J.~Wess and B.~Zumino,
  Phys.\ Rev.\  {\bf 177}, 2239 (1969);
  C.~G.~Callan, Jr., S.~R.~Coleman, J.~Wess and B.~Zumino,
  Phys.\ Rev.\  {\bf 177}, 2247 (1969).

\bibitem{eftorig}
  C.~J.~C.~Burges and H.~J.~Schnitzer,
  Nucl.\ Phys.\ B {\bf 228}, 464 (1983); 
  C.~N.~Leung, S.~T.~Love and S.~Rao,
  Z.\ Phys.\ C {\bf 31}, 433 (1986); 
  W.~Buchm\"uller and D.~Wyler,
  Nucl.\ Phys.\ B {\bf 268}, 621 (1986);
  W.~Kilian,
  Springer Tracts Mod.\ Phys.\  {\bf 198}, 1 (2003).
  
\bibitem{higgsreview}
  C.~Englert, A.~Freitas, M.~M.~M\"uhlleitner, T.~Plehn, M.~Rauch, M.~Spira and K.~Walz,
  J.\ Phys.\ G {\bf 41}, 113001 (2014).

\bibitem{legacy}
  T.~Corbett, O.~J.~P.~\'Eboli, D.~Gon\c{c}alves, J.~Gonzalez-Fraile, T.~Plehn and M.~Rauch,
  JHEP {\bf 1508}, 156 (2015).
  
\bibitem{heft_limitations} 
  C.~Arnesen, I.~Z.~Rothstein and J.~Zupan,
  Phys.\ Rev.\ Lett.\  {\bf 103}, 151801 (2009);
  C.~Englert and M.~Spannowsky,
  Phys.\ Lett.\ B {\bf 740}, 8 (2015);
  M.~de Vries,
  JHEP {\bf 1503}, 095 (2015);
  N.~Craig, M.~Farina, M.~McCullough and M.~Perelstein,
  JHEP {\bf 1503}, 146 (2015);
  S.~Dawson, I.~M.~Lewis and M.~Zeng,
  Phys.\ Rev.\ D {\bf 91}, 074012 (2015);
  A.~Drozd, J.~Ellis, J.~Quevillon and T.~You,
  JHEP {\bf 1506}, 028 (2015);
  R.~Edezhath,
  arXiv:1501.00992 [hep-ph];
  L.~Edelh\"auser, A.~Knochel and T.~Steeger,
  arXiv:1503.05078 [hep-ph];
  M.~Gorbahn, J.~M.~No and V.~Sanz,
  arXiv:1502.07352 [hep-ph];

\bibitem{too_long} 
  J.~Brehmer, A.~Freitas, D.~Lopez-Val and T.~Plehn,
  arXiv:1510.03443 [hep-ph].

\bibitem{mvh} 
  J.~Ellis, V.~Sanz and T.~You,
  JHEP {\bf 1503}, 157 (2015).

\bibitem{gino}
  A.~Greljo, G.~Isidori, J.~M.~Lindert and D.~Marzocca,
  arXiv:1512.06135 [hep-ph].

\bibitem{square_dim_6_others} 
  L.~Berthier and M.~Trott,
  JHEP {\bf 1505}, 024 (2015);
  L.~Berthier and M.~Trott,
  JHEP {\bf 1602}, 069 (2016);
  R.~Contino, A.~Falkowski, F.~Goertz, C.~Grojean and F.~Riva,
  JHEP {\bf 1607}, 144 (2016);
  O.~Bessidskaia Bylund, F.~Maltoni, I.~Tsinikos, E.~Vryonidou and C.~Zhang,
  JHEP {\bf 1605}, 052 (2016);
  F.~Maltoni, E.~Vryonidou and C.~Zhang,
  arXiv:1607.05330 [hep-ph].

\bibitem{spanno}
  for a different point of view see \eg 
  C.~Englert, R.~Kogler, H.~Schulz and M.~Spannowsky,
  arXiv:1511.05170 [hep-ph].

\bibitem{gauge_modifications} 
  I.~Low, R.~Rattazzi and A.~Vichi,
  JHEP {\bf 1004}, 126 (2010);
  D.~Pappadopulo, A.~Thamm, R.~Torre and A.~Wulzer,
  JHEP {\bf 1409}, 060 (2014); 
  A.~Thamm, R.~Torre and A.~Wulzer,
  Phys.\ Rev.\ Lett.\  {\bf 115}, no. 22, 221802 (2015);
  a FeynRules implementation
  of the model is available from 
  \texttt{http://heidi.pd.infn.it/html/vector/index.html}

\bibitem{anke} 
  J.~de Blas, J.~M.~Lizana and M.~Perez-Victoria,
  JHEP {\bf 1301}, 166 (2013);
  A.~Biekoetter, A.~Knochel, M.~Kr\"amer, D.~Liu and F.~Riva,
  Phys.\ Rev.\ D {\bf 91}, 055029 (2015);
  A.~E.~C.~Hern\'andez, C.~O.~Dib and A.~R.~Zerwekh,
  arXiv:1506.03631 [hep-ph].

\bibitem{hisz} 
  K.~Hagiwara, S.~Ishihara, R.~Szalapski and D.~Zeppenfeld,
  Phys.\ Rev.\ D {\bf 48}, 2182 (1993).
  
\bibitem{hh}
  U.~Baur, T.~Plehn and D.~L.~Rainwater,
  Phys.\ Rev.\ D {\bf 69}, 053004 (2004).
 
\bibitem{madgraph}
  J.~Alwall {\it et al.},
  JHEP {\bf 1407}, 079 (2014).

\bibitem{kilian}
  W.~Kilian, T.~Ohl, J.~Reuter and M.~Sekulla,
  Phys.\ Rev.\ D {\bf 91}, 096007 (2015).

\bibitem{effective_w}
  S.~Dawson,
  Nucl.\ Phys.\ B {\bf 249}, 42 (1985);
  G.~L.~Kane, W.~W.~Repko and W.~B.~Rolnick,
  Phys.\ Lett.\ B {\bf 148}, 367 (1984);
  J.~Alwall, D.~Rainwater and T.~Plehn,
  Phys.\ Rev.\ D {\bf 76}, 055006 (2007);
  P.~Borel, R.~Franceschini, R.~Rattazzi and A.~Wulzer,
  JHEP {\bf 1206}, 122 (2012).

\bibitem{polarized_ww}
  J.~Brehmer, J.~Jaeckel and T.~Plehn,
  Phys.\ Rev.\ D {\bf 90}, no. 5, 054023 (2014).
  
\bibitem{Buschmann:2014twa} 
  M.~Buschmann, C.~Englert, D.~Gon\c{c}alves, T.~Plehn and M.~Spannowsky,
  Phys.\ Rev.\ D {\bf 90}, no. 1, 013010 (2014).

\bibitem{spins1}
  K.~Hagiwara, Q.~Li and K.~Mawatari,
  JHEP {\bf 0907}, 101 (2009);

\bibitem{spins2}
  C.~Englert, D.~Gon\c{c}alves-Netto, K.~Mawatari and T.~Plehn,
  JHEP {\bf 1301}, 148 (2013);
  A.~Djouadi, R.~M.~Godbole, B.~Mellado and K.~Mohan,
  Phys.\ Lett.\ B {\bf 723}, 307 (2013).

\bibitem{pythia}
  T.~Sjostrand, S.~Mrenna and P.~Z.~Skands,
  JHEP {\bf 0605}, 026 (2006).

\bibitem{mlm}
  M.~L.~Mangano, M.~Moretti, R.~Pittau,
  Nucl.\ Phys.\  {\bf B632}, 343-362 (2002).

\bibitem{fastjet}
  M.~Cacciari, G.~P.~Salam and G.~Soyez,
  JHEP {\bf 0804}, 063 (2008);
  M.~Cacciari, G.~P.~Salam and G.~Soyez,
  Eur.\ Phys.\ J.\ C {\bf 72}, 1896 (2012).

\bibitem{simp} 
  D.~Alves {\it et al.} [LHC New Physics Working Group Collaboration],
  J.\ Phys.\ G {\bf 39}, 105005 (2012).

\bibitem{simp_higgs} 
  M.~J.~Dolan, J.~L.~Hewett, M.~Kr\"{a}mer and T.~G.~Rizzo,
  arXiv:1601.07208 [hep-ph].

\bibitem{delta_phi} 
  O.~J.~P.~Eboli and D.~Zeppenfeld,
  Phys.\ Lett.\ B {\bf 495}, 147 (2000);
  T.~Plehn, D.~L.~Rainwater and D.~Zeppenfeld,
  Phys.\ Rev.\ Lett.\  {\bf 88}, 051801 (2002);
  V.~Hankele, G.~Klamke, D.~Zeppenfeld and T.~Figy,
  Phys.\ Rev.\ D {\bf 74}, 095001 (2006);
  M.~R.~Buckley, T.~Plehn and M.~J.~Ramsey-Musolf,
  Phys.\ Rev.\ D {\bf 90}, no. 1, 014046 (2014).

\bibitem{effective_scalar}
  J.~R.~Ellis, M.~K.~Gaillard and D.~V.~Nanopoulos,
  Nucl.\ Phys.\ B {\bf 106}, 292 (1976);
  S.~Dawson and L.~Reina,
  Phys.\ Rev.\ D {\bf 57}, 5851 (1998);
  E.~Braaten and H.~Zhang,
  arXiv:1510.01686 [hep-ph].

\end{thebibliography}
